\documentclass[aps,twocolumn,amsmath,amssymb,superscriptaddress]{revtex4}

\usepackage{graphicx}
\usepackage{dcolumn}
\usepackage{bm}
\usepackage{psfrag}

\DeclareMathOperator{\sgn}{sgn}
\DeclareMathOperator{\acosh}{acosh}
\DeclareMathOperator{\acos}{acos}
\DeclareMathOperator{\Li}{Li}
\DeclareMathOperator{\Tr}{Tr}

\def\Xint#1{\mathchoice
   {\XXint\displaystyle\textstyle{#1}}%
   {\XXint\textstyle\scriptstyle{#1}}%
   {\XXint\scriptstyle\scriptscriptstyle{#1}}%
   {\XXint\scriptscriptstyle\scriptscriptstyle{#1}}%
   \!\int}
\def\XXint#1#2#3{{\setbox0=\hbox{$#1{#2#3}{\int}$}
     \vcenter{\hbox{$#2#3$}}\kern-.5\wd0}}

\def\dashint{\Xint-}

\newcommand{\be}{\begin{equation}}
\newcommand{\ee}{\end{equation}}
\newcommand{\bea}{\begin{eqnarray}}
\newcommand{\eea}{\end{eqnarray}}
\newcommand{\HH}{{\cal H}}

\newcommand{\tr}{{\rm tr\/}\,}
\newcommand{\p}{\partial}

\newcommand{\la}{\langle}
\newcommand{\ra}{\rangle}
\newcommand{\lb}{\left[}
\newcommand{\rb}{\right]}
\newcommand{\lp}{\left(}
\newcommand{\rp}{\right)}
\renewcommand{\phi}{\varphi}
\renewcommand{\vec}[1]{{\bf #1}}
\renewcommand{\Re}{{\rm Re}\,}
\renewcommand{\Im}{{\rm Im}\,}

\numberwithin{equation}{section}
\renewcommand{\theequation}{\arabic{section}.\arabic{equation}}
\begin{document}

\title{Entropy and Correlation Functions of a Driven Quantum Spin Chain}
%% Nonequilbrium Dynamics, Decoherence, and Correlations of Quantum Spin Chains}
\author         {R. W. Cherng}
\affiliation{Department of Physics, Harvard University, Cambridge, Massachusetts 02138, USA}
\author         {L. S. Levitov}
\affiliation{Department of Physics, Massachusetts Institute of Technology, 77 Massachusetts Ave,
Cambridge, Massachusetts 02139, USA}
\date{\today} 

\begin{abstract}
We present an exact solution for a quantum spin chain driven 
through its critical points. Our approach is based on a many-body
generalization of the Landau-Zener transition theory,
applied to fermionized spin Hamiltonian.
The resulting nonequilibrium state of the system, 
% evolved from the initial ground state, 
while
being a pure quantum state,
has local properties of a mixed state
characterized by finite entropy density
associated with Kibble-Zurek defects.
The entropy, as well as the finite spin correlation length,
are functions of the rate of sweep through the critical point.
We analyze the 
anisotropic XY spin $1/2$ model evolved with a full many-body evolution operator.
With the help of Toeplitz determinants
calculus, we obtain an exact form of correlation functions.
The properties of the evolved system undergo an abrupt 
change at a certain critical sweep rate, signaling formation of ordered
domains. We link this phenomenon to the
behavior of complex singularities of the Toeplitz generating function.
\end{abstract}

\maketitle

\section{Introduction}
\label{sec:introduction}
Recent advances in the studies
of ultracold atoms trapped in optical lattices have opened a new arena
of investigation of nonequilibrium
strongly correlated quantum systems \cite{greiner1,jaksch}.
These new opportunities are epitomized by the pioneering experiments on tunable
Mott insulator-to-superfluid quantum phase transition, observed by
manipulation of the optical lattice potential 
in 3d \cite{greiner1} and 1d \cite{stoferle} systems.
The highly controllable environment and long coherence times of these systems 
provide new framework for investigation of nonequilibrium
dynamics of quantum critical phenomena \cite{clark,zurek1,sachdev1}.

One interesting question arising in this framework
has to do with the properties 
of defects produced by
sweeping through a critical point. For the phase transitions
occurring at finite temperature the defect production
is described by Kibble-Zurek (KZ) theory \cite{kibble,zurek2}.
This theory, which initially was applied to topological
defects left behind cosmological phase transitions,
and only later found its way in condensed matter physics,
estimates the correlation length 
in the ordered state using a causality argument.
The correlation length serves as a measure of the size of the ordered domains
and of typical separation between defects. 
Defect production was probed in
recent experiments employing superfluid ${}^3{\rm He}$ \cite{ruutu96,bauerle96}
and superconducting 
Josephson junctions \cite{carmi00}.

Phase transitions in cold atom systems 
are characterized by a high degree of coherence, which
makes the dynamics near the critical point essentially non-dissipative.
The theory of defect production in this situation has to be modified
to account for coherent dynamics.
Defect production in quantum dynamics can be studied
using integrable 1d spin models. 
The 1d spin models
with varying coupling constants
provide a template for many quantum phenomena.
Realizations of such models
have been proposed recently in 1d qubit chains\cite{levitov01}
and optical lattices\cite{demler}.
The models of quantum spin quench dynamics
resulting from an abrupt change of coupling constant which takes the
system across the phase boundary, were considered
in Refs.\cite{sachdev1,Cardy05}. 
The quench dynamics, while providing useful insight,
do not describe the situation of
a continuous sweep across the transition,
which is addressed in the present work.

Besides defect production rate and density,
there is an interesting question of the entropy associated with 
the defects. Naively, it may seem that the entropy cannot be produced
at zero temperature by a system evolving unitarily in a pure state.
However, if the evolved state is sufficiently complex, it may look
entropic from a local point of view, i.e. if observed in a volume 
much smaller than the total system size. 
As we shall see, this is precisely
the case in this problem.

%% A recent proposal \cite{demler} shows how many-body quantum spin systems  
%% can be engineered in optically trapped ultracold atoms.
%% One-dimensional quantum spin chains are the simplest spin systems and are the
%% prototypical example of quantum criticality  \cite{sachdev3}.
%% The dynamics of spin chains even in highly nonequilibrium regimes 
%% is still amenable to analytical and numerical methods.
%% This includes aspects of nonequilibrium dynamics including quantum 
%% aging \cite{igloi}, generation
%% of maximally entangled states \cite{dorner}, 
%% and nonequilibrium steady states \cite{antal}.

In the present article we study time evolution of 
a many-body system which is swept 
at a constant speed through its quantum critical point. 
With the help of an exactly solvable 1d 
quantum spin model with a time-dependent Hamiltonian we explore
how the time evolution across the critical point manifests itself
in the many-body effects and spin correlation functions.
In particular, we analyze
the relation between the sweep speed 
and spatial spin correlations, providing an extension of  
KZ scenario to the quantum critical point regime.
Our analytical results are in agreement with recent numerical study
of this problem, reported in Ref.\cite{zurek5}.

Our approach is based on a many-body generalization of the
Landau-Zener (LZ) transition theory.
In this work we focus
on the anisotropic XY spin $1/2$ chain with
time-dependent couplings.
We consider unitary evolution of the system, initially
in the ground state, which 
crosses its equilibrium critical points. 
Since the Hamiltonian of the fermionized spin chain is quadratic,
the evolution of the many-body state can be expressed
with the help of a Bogoliubov
transformation through
a suitable set of the $2\times2$  
evolution problems of LZ form, one for each fermion momentum value.

Our analysis reveals that
the evolved system state
has a number of interesting characteristics.
Firstly, despite being in a pure quantum state in a global sense,
its local properties are identical to those
of a system in a mixed state, characterized 
by finite effective temperature and entropy density. 
Although the finite entropy property
of a pure state may seem counterintuitive, it naturally
arises in the description of local properties, such as correlation
functions.
We shall see that the origin of finite entropy
can be traced to coarse-graining in momentum space.
On a more intuitive level, the system pure state can described 
as a superposition of different configurations of ordered domains
with uniform magnetization. However, the coherence of amplitudes
associated with different domain arrangements cannot be detected 
locally without having access to the entire set of variables in the system,
which leads to an apparent mixed state and finite entropy.

Secondly, the transition from the adiabatic
to non-adiabatic regime in the LZ problem, 
taken as a function of the sweep rate,
depends on the momentum value of the fermionic mode. The characteristic
crossover momentum can be associated with the inverse correlation length
$\ell$
in the KZ picture, corresponding to typical domain size.
This approach yields a scaling relation between
the correlation length and the sweep speed, $\ell\propto v^{-1/2}$.
This relation, obtained directly from the analysis of the many-body evolution 
operator, agrees with the KZ causality argument prediction.

\begin{figure}
\begin{center}
\includegraphics[width=3.4in]{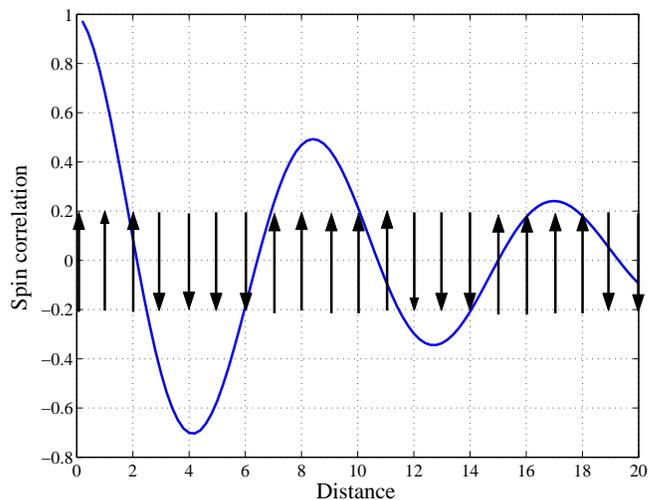}
\caption{
Spin correlation function schematic position dependence
for slow sweep speed
and corresponding typical arrangement of Kibble-Zurek domains.
The correlation length and
oscillation period are controlled by domain size.
}
\label{fig:correlator_oscill} 
\end{center} 
\end{figure}

Lastly, due to a simple product structure of the evolved state, the correlation functions
can be obtained in a closed, exact form with the help 
of the theory of Toeplitz determinants.
The correlation functions exhibit 
a crossover from monotonically decreasing behavior at fast sweep speed, $e^{-r/\ell}$,
to an oscillatory behavior at a slow speed, $e^{-r/\ell}\cos(\omega r-\phi)$. 
The oscillatory behavior, which appears abruptly below certain
sweep speed value, corresponds to alternate magnetization 
signs in neighboring ordered domains (see Fig. \ref{fig:correlator_oscill}). The spatial period
$2\pi/\omega$ gives characteristic domain size.
The parameters $\ell$, $\omega$ and $\phi$ exhibit 
a singularity 
at the critical sweep speed, which is analyzed 
and explained in the Toeplitz determinant framework
via evolution of zeroes of the generating function in a complex plane.

The plan of this article is as follows.
We start with analyzing the full many-body evolution operator 
of the XY spin chain with the help of Jordan-Wigner
fermionization and reduction to the LZ transition problem
in each fermion momentum subspace
(Sec. \ref{sec:dynamics}).  
Next, in Sec. \ref{sec:decoherence}, we show that
in a macroscopic system
(number of sites $N\to\infty$), a non-equilibrium steady state (NESS) 
emerges at late times. This is a mixed state characterized by a density matrix
with finite entropy which depends on the sweep speed.
The state of a mixed character appears due to decoherence intrinsic to the many-body LZ process,
without any external decoherence effects. 
Technically, the mixed state 
arises as a result of taking the large $N$ limit in the 
correlation functions for spins separated by distances much less than
the system size, $r\ll N$. 
This procedure allows to eliminate the rapidly oscillating terms 
in the correlation functions, which would
disappear in a real system as a result of physical decoherence processes, 
even if the latter are extremely weak.
The entropy of NESS
is analyzed in Sec. \ref{sec:entropy_new}.

The density matrix description of NESS is subsequently used 
in Secs. \ref{sec:correlators1} and \ref{sec:correlators2} to characterize ordering and analyze correlation functions. 
The method employed in analytic calculation uses some
results from the theory of Toeplitz determinants which are reviewed in Appendix A.
We obtain the asymptotics of equal-time spin correlators in the NESS 
which have non-trivial crossover behavior as a function
of the sweep rate. Both numerical and analytical results are presented, compared,
and found to be in agreement.

\section{Spin Chain Dynamics}
\label{sec:dynamics}

In this section, we 
consider 
a quantum XY spin $1/2$ chain in time-dependent transverse field, 
described by the Hamiltonian
\begin{eqnarray}
\label{eq:hamiltonian}
\nonumber
\HH(t)&=&-\frac{1}{2}\sum_{x=1}^N\left[J_1\sigma_x^1\sigma_{x+1}^1+J_2\sigma_x^2\sigma_{x+1}^2-h(t)\sigma_x^3\right]
\end{eqnarray}
where $N$ is the number of sites. The anisotropic coupling values are
\be
J_1=J(1+\gamma)/2,\quad
J_2=J(1-\gamma)/2,\quad
h(t)=vt
\ee
Here $J=\frac12(J_1+J_2)$ is the average coupling and 
$\gamma=(J_1-J_2)/(J_1+J_2)$ is the anisotropy parameter. Note 
that the values $\gamma=0,\pm 1$ describe the isotropic XY model 
and the Ising model, respectively. (Without loss of generality,
we assume  $J>0$.) 

In this article,
the problem (\ref{eq:hamiltonian}) is considered with periodic
boundary conditions, i.e. $x=N+1$ is identified with $x=1$.
Other choices, such as open boundary conditions, are possible.
While the properties of interest in the large $N$ limit 
will be insensitive to the form of boundary conditions, 
periodic boundary conditions will make the intermediate steps
of calculations more transparent.

\begin{figure}
\begin{center}
\includegraphics[width=2.0in]{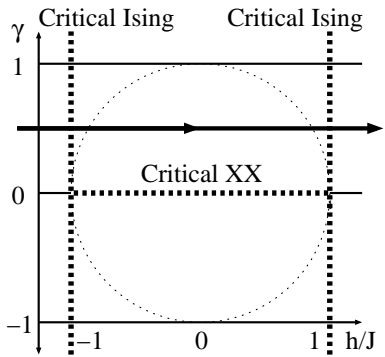}
\caption{
Zero-temperature phase diagram of anisotropic XY model 
adapted from Ref.\cite{barouch2}.  
The lines of critical points,
$h=\pm J$ and $\gamma=0$ with $|h/J|\le1$ are marked by dashed lines, 
the circular domain $(h/J)^2+\gamma^2\le 1$ is marked by dotted line
(see text). 
The evolution trajectory of the system
(\ref{eq:hamiltonian}) due to time-dependent 
$h(t)$ is shown by solid line.}
\label{fig:phase_diagram} 
\end{center} 
\end{figure}

The time-dependent transverse field $h(t)$ defines 
the evolution in the equilibrium system phase space which starts from and ends at the 
state in which the external field $h(t)$ is much larger 
than the couplings $J_{1,2}$ (Fig.\ref{fig:phase_diagram}). 
Thus in the asymptotic ground states at $t\to\pm\infty$ the spins are fully polarized: 
$\psi^{(0)}_{-\infty}=(...\downarrow\downarrow\downarrow\downarrow...)$ and
$\psi^{(0)}_{+\infty}=(...\uparrow\uparrow\uparrow\uparrow...)$.
A fully adiabatic time evolution (with negligible speed $v=dh/dt$) 
would transform
the initial state $\psi^{(0)}_{-\infty}$
into the state $\psi^{(0)}_{+\infty}$.
This would also describe physical evolution 
at a finite but sufficiently slow speed,
provided that the ground and excited states
are separated by a finite gap at all times.
However, if the evolution takes the system through 
a critical point, where the gap vanishes, the nonadiabatic effects 
inevitably give rise to a state much more complex than 
$(...\uparrow\uparrow\uparrow\uparrow...)$.

To analyze the time-dependent state
we evaluate the evolution operator 
$\hat U_T={\rm Texp}\lp -i\int_{-T}^{T}H(t)dt\rp$, 
using Schr\"odinger representation. 
We choose  a long evolution time interval,
$-T<t< T$, so that
\be\label{eq:T>>tQ}
T\gg t_Q\equiv J/v
,
\ee
where $2t_Q$ is the transit time between the 
critical lines $h=\pm J$ (Fig. \ref{fig:phase_diagram}).
Since the effect of the couplings $J_{1,2}$ is important only during 
a relatively short
time interval of order $t_Q$, when $h(t)\simeq J_{1,2}$, one expects the results 
to be fairly insensitive to the specific value of $T$.
Indeed, as we shall discover shortly, in the limit described by Eq.(\ref{eq:T>>tQ})
universal results will arise.

The model (\ref{eq:hamiltonian})
has a long history dating back to the original
solution of the equilibrium model by Lieb, Schulz, and Mattis \cite{lsm}
who obtained an exact solution using Jordan-Wigner fermionization. 
Let us recall the basic features of the phase diagram in equilibrium. 
Barouch and McCoy
\cite{barouch2} obtained  
the phase diagram 
by considering spin correlators in the ground state. These results
were subsequently extended by Tracy and Vaidya \cite{vaidya2,vaidya1}
and further generalized in Refs.\cite{korepin2,korepin1} 
which employ quantum inverse scattering technique.

For reader's convenience, here we summarize 
the zero-temperature equilibrium phase diagram \cite{barouch2} 
as a function of $h/J$ and $\gamma$ 
in Fig. \ref{fig:phase_diagram}. 
The system exhibits spontaneous ferromagnetic Ising order 
for $-J<h<J$, (antiferromagnetic for $J<0$)
and can be described for $|h|>J$ as 
disordered, or paramagnetic.  The lines of critical points $h=\pm J$, 
separating these regimes, are in the Ising universality class.
The gap in the excitation spectrum 
\be\label{eq:epsilon(k)}
\epsilon(k)=\pm \lp (h+J\cos k)^2+\gamma^2 J^2\sin^2 k\rp^{1/2}
\ee
vanishes on the critical lines.
Outside the circular domain marked in Fig. \ref{fig:phase_diagram}, 
$\gamma^2+h^2/J^2>1$, the correlators in the ground state
exhibit Ising-like pure exponential decay.
In contrast, for $\gamma^2+h^2/J^2\le1$
the correlators have
oscillatory subleading terms.
The ground state on the circle $\gamma^2+h^2/J^2=1$ is 
a direct product of single-site spin states \cite{muller}.
On the $\gamma=0$ line ($J_1=J_2$) the Hamiltonian is isotropic.  
In this case, in the interval $-J<h<J$
the ground state is quantum critical.

For our choice of the time-dependent field, the system is deep 
in the disordered phase at both
the early and late times, $|h(t\sim\pm T)|\gg J$.  
At such times the instantaneous eigenstates of $H(t)$ 
evolve quasi-adiabatically, 
with a pure phase factor. However, at intermediate times $t\simeq t_Q$ 
we expect non-trivial dynamics as the system enters the phase with 
spontaneous Ising order, $-J<h<J$,  
passing through the critical points at $h(t)=\pm J$.

Our exact solution of the dynamical problem 
is a direct generalization
of the equilibrium solution.
We employ the time-independent Jordan-Wigner string variables
\be
\label{eq:strings}
\tau_x=\prod_{x'<x}(-\sigma_{x'}^3).
\ee
In the Ising limit $\gamma=1$, the quantities $\tau_x$ are dual to 
the $\sigma_x^1$ and
represent so-called disorder variables \cite{savit}.
With the help of $\tau_x$ we define spinless fermionic operators
%
%\label{eq:jw_fermion}
\[
a_x=\tau_x\sigma_x^-,\quad
a_x^+=\tau_x\sigma_x^+,
\]
%\ee
with $\sigma_x^{\pm}=\frac12(\sigma_x^1\pm i\sigma_x^2)$ 
the raising and lowering operators.

The fermionized Hamiltonian is quadratic:
\be\label{eq:Hfermion}
\HH=\sum_{x=1}^N 
A_xa^+_xa_{x+1}+B_xa_xa_{x+1}+{\rm h.c.}-2h(t)a^+_xa_x
,
\ee
where we subtracted a constant $E_0=Nh(t)$.
Here the couplings $A_x=J_1+J_2=J$, $B_x=J_2-J_1=-\gamma J$ 
are the same for all $1\le x<N$, and
\be\label{eq:tauN}
A_{x=N}=J\tau_N,\quad B_{x=N}=-\gamma J\tau_N.
\ee
The string operator $\tau_N$ can be expressed as $\exp(i\pi\hat {\cal N})$,
where $\hat {\cal N}=\sum_{x=1}^N a^+_xa_x$ is the total fermion number.
The complication due to the presence of the operator-valued
couplings (\ref{eq:tauN}) in the Hamiltonian (\ref{eq:Hfermion})
turns out to be inessential \cite{barouch2}. In fact, since different
terms of Eq.(\ref{eq:Hfermion}) either conserve the fermion number $\hat {\cal N}$,
or change it by $\pm 2$, the operator $\tau_N$ is a constant
of motion, $[\tau_N,\HH]=0$. This allows to replace $\tau_N$ by the c-number
equal to its value in the initial state:
$\tau_N=(-1)^N$. Thus we obtain a truly quadratic translationally invariant 
Hamiltonian in the fermion representation 
with periodic or antiperiodic boundary 
conditions, depending on the parity of $N$.

It will be convenient to write fermionic operators using two-component 
vectors, 
\be
\label{eq:vector_fermions}
\vec{C}_x=
\lp \begin{matrix}
a_x\\
a^\dagger_x
\end{matrix}\rp,\ \ \
\vec{C}_k=
\lp \begin{matrix}
a_k\\
a^\dagger_{-k}
\end{matrix}\rp 
= \frac{1}{\sqrt{N}}\sum_x e^{ikx}\vec{C}_x,
\ee
with $k=2\pi m/N$, where $m$ is integer or half-integer, 
depending on the parity of $N$.
The fermionized Hamiltonian, in the momentum representation
(\ref{eq:vector_fermions}),
splits into a sum of independent terms,
$H(t)=\lp \sum_{k>0}H_k(t)\rp +E'_0$, where 
each term operates in the four-dimensional Hilbert space
associated with the momentum states $k$, $-k$ 
filled with different numbers of fermions, 
and $E'_0=\sum_{k\ge0}J\cos k$ is a constant.
The operators $H_k(t)$ are bilinear in $\vec{C}_k$ and have the form
\be
\label{eq:fermion_hamiltonian}
H_k(t)=-\vec{C}_k^{\dagger}\left(
\begin{matrix}
h(t)+J\cos k & i\gamma J\sin k\\ -i\gamma J\sin k & -h(t)-J\cos k
\end{matrix}
\right)\vec{C}_k
\ee
which conserves $k$ due to translational invariance.
Also, $H_k$ conserves
the fermion occupancy number $n_k=a^+_ka_k+a^+_{-k}a_{-k}$ up to $\pm2$ 
(i.e. the parity of $n_k$) 
separately within each $k$-subspace $(k,-k)$.

\section{Many-body Landau-Zener transition}
\label{sec:LZtransition}

Using the representation (\ref{eq:fermion_hamiltonian})
we can write the full many-body evolution operator as a tensor product
of partial evolution operators acting in the $(k,-k)$ subspaces:
\be
U(t)=\bigotimes_{k>0}\hat U_k(t)
\,,\quad
\hat U_k(t) = {\rm Texp}\lp -i\int_{-T}^{t}H_k(t')dt'\rp
\ee
To obtain $\hat U_k$, we consider the basis 
in the $k$, $-k$ subspace generated by
the $a_{k}$ vacuum $a_k|0\rangle=0$ as follows:
\begin{eqnarray}
\label{eq:ak_basis}
\nonumber
&& |0\rangle,\quad |k,-k\rangle = a^\dagger_k a^\dagger_{-k}|0\rangle\\
\nonumber
&& |k\rangle = a^\dagger_k|0\rangle.\quad |-k\rangle = a^\dagger_{-k}|0\rangle
\nonumber
\end{eqnarray}
The latter two states $|\pm k\rangle$ of occupancy one 
are eigenstates of the Hamiltonian
(\ref{eq:fermion_hamiltonian}): 
\[
H_k(t)|\pm k\rangle=(h(t)+J\cos k)|\pm k\rangle
.
\]
(This follows from
conservation of $k$ and 
the parity of $n_k$.)
Thus each of the states $|\pm k\rangle$ evolves in time with a phase factor,
$|\pm k\rangle(t)=e^{-i\phi(t)}|\pm k\rangle$, with
\be
\frac{d\phi}{dt}=h(t)+J\cos k
.
\ee
The other two states, $|0\rangle$ and $|k,-k\rangle$,
evolve as superposition 
$\Psi_k(t)=u_k(t)|0\rangle+v_k(t)|k,-k\rangle$. 
We denote the corresponding $2\times2$ evolution operator as $\hat S_k(t)$.

This discussion can be summarized by writing 
the $4\times4$ evolution operator $\hat U_k$ in a block-diagonal form:
\be\label{eq:Uk4times4}
\hat U_k=\left(
\begin{matrix}
\hat S_k(t) & 0\\ 0 & e^{-i\phi(t)\hat 1}
\end{matrix}
\right)
\ee
with $\hat 1$ a $2\times2$ identity operator.
The first and the second block
correspond to the states $|0\rangle$, $|k,-k\rangle$
and $|\pm k\rangle$, respectively.

To describe $\hat S_k(t)$,
we project the Hamiltonian $H_k(t)$ on the subspace
$|0\rangle$, $|k,-k\rangle$, which gives
an evolution equation for $u_k(t)$, $v_k(t)$ as follows:
\be\label{eq:two_state}
i\p_t\Psi_k=
\left(
\begin{matrix}
h(t)+J\cos k & -2i\gamma J\sin k\\ 2i\gamma J\sin k & -h(t)-J\cos k
\end{matrix}
\right)\Psi_k
\ee
The form of Eq.(\ref{eq:two_state}) is identical
to that of the LZ transition problem \cite{landau,zener} for
two levels evolving linearly with time through an avoided crossing 
of size $\Delta_k=2\gamma J|\sin k|$.

The result of the evolution defined by Eq.(\ref{eq:two_state})
can be represented as a $2\times2$ unitary matrix
which depends on the Landau-Zener adiabaticity parameter
$\alpha_k=|\Delta_k|^2/v_{12}$, where $v_{12}=2dh/dt=2v$
is the relative velocity of the levels.
The parameter
$\alpha_k$ is small for fast level crossing and large for slow crossing.
In our case, we have
\[
\alpha_k=(4\gamma^2J^2/2v)\sin^2k \equiv z\sin^2k,
\] 
where we introduced the dimensionless parameter
\be
\label{eq:LZ_parameter}
z=2\gamma^2 J^2/v
\ee
to be used throughout the rest of the paper.

The evolution matrix for the LZ problem can be obtained exactly 
in analytic form.
In the limit of the total evolution time long compared to
the level crossing time 
(realized in our case, since $T/t_Q\gg 1$),
one can write the evolution operator $S_k$ explicitly
in terms of the LZ transiton amplitudes
\be\label{eq:asymptotic_r,s}
r_k = e^{-\pi\alpha_k} 
,\quad
s_k = -\sgn(k)\sqrt{1-r_k^2}
\ee
The long-time asymptotic 
form of the matrix $\hat S_k$ (e.g., see Ref.\cite{brundobler}) 
is as follows:
\be
\label{eq:fermion_smatrix}
\nonumber
S_k=
\begin{bmatrix}
r_k e^{-i\phi_k}&   -s_k e^{-i\eta_k} \\
s_k e^{i\eta_k}& r_k e^{i\phi_k}
\end{bmatrix},
\ee
where the time-dependent phases are 
\begin{eqnarray}
\label{eq:asymptotic_smatrix}
\phi_k&=&f_k(x_k^+)-f_k(x_k^-)\\\nonumber
\eta_k&=&f_k(x_k^+)+f_k(x_k^-)+\pi/4-\arg\Gamma(i\alpha_k)\\\nonumber
f_k(x)&=&x^2/4+\alpha_k\ln|x|+\mathcal{O}(x^{-2})
\end{eqnarray}
Here 
$\Gamma(x)$ is 
the gamma function and
\be
x_k(t)=2(vt+J\cos k)/v^{1/2}
\ee
is dimensionless time.
Note that in the long time limit only the phases $\phi_k$, $\eta_k$ depend
on time, quickly growing as a function of $T$, while 
the amplitudes $r_k$, $s_k$ become time-independent,
approaching the asymptotic values (\ref{eq:asymptotic_r,s}).

\begin{figure}
\begin{center}
\includegraphics[width=8.5cm]{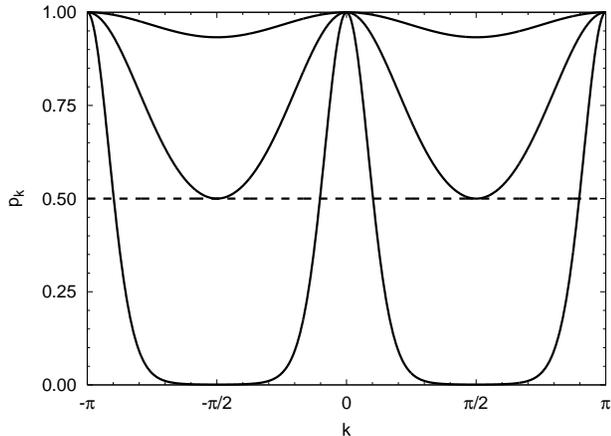}
\caption{
LZ probability (\ref{eq:LZprob}) of remaining in the initial state
for $z/z_\ast=0.1,1,10$ (from top to bottom).
The dashed
line marks $p_k=0.5$. 
Note the regions near $k=0,\pm\pi$ (critical modes) 
where LZ transition does not take place
even at a slow sweep speed $z/z_\ast\gg1$.
}
\label{fig:pk} 
\end{center} 
\end{figure}

Since the states $0\rangle$, $|\pm k\rangle$, $|k,-k\rangle$ 
are invariant (up to a phase factor) 
at $t\to\pm\infty$, with LZ transitions between
$0\rangle$ and $|k,-k\rangle$ happening only at times $t\simeq J/v$,
the asymptotic matrix $S_k$ can be used to describe transitions
resulting from the time evolution. 
In Fig. \ref{fig:pk}
we plot the probability 
\be\label{eq:LZprob}
p_k=|r_k|^2=e^{-2\pi\alpha_k}=e^{-2\pi z\sin^2k}
\ee
for the system, evolving 
from the state
$|0\rangle$ at $t=-T$, to remain
in this state at late time $t=T$. (The quantity (\ref{eq:LZprob})
also describes
the probability of the state $|k,-k\rangle$ to remain itself.) 
The top curve in Fig. \ref{fig:pk} 
corresponds to small $z$ (fast sweep rate $v$)
when the levels cross quickly and the transition
probability is small. 
The transition probability increases at larger $z$, with fully adiabatic regime
reached for typical values of $k$ at very large $z$.
In this limit, the systems performs a nearly complete 
transfer of population from the initial state $|0\rangle$
to the state $|k,-k\rangle$, which in the spin
language corresponds to spin orientation reversal $\sigma^3_x\to -\sigma^3_x$.
This behavior is illustrated by the lower curve in Fig. \ref{fig:pk}.
In this case, while the majority of the modes evolve 
adiabatically to the final state $|k,-k\ra$,
a small fraction of the modes 
with $k$ close to $0,\pm\pi$ evolve nonadiabatically. 
These modes remain stuck
in the the initial state $|0\ra$, for $p_k\approx 1$, or form
a superposition of the states $|0\ra$
and $|k,-k\ra$ with comparable weights, for $p_k\approx 1/2$ (see Fig. \ref{fig:pk}).

To characterize the degree of adiabaticity of different modes, it is convenient
to define a special 
value of $z$ which will be of importance in the discussion below:
\be
\label{eq:zstar}
z_\ast=\frac{\ln 2}{2 \pi}= 0.110...
\ee
As Fig. \ref{fig:pk} illustrates, at $z=z_\ast$ the curve $p_k$ is tangent
to the $p=1/2$ line at $k=\pm\pi/2$. 
As we shall see 
in Sec.\ref{sec:decoherence}, the modes with $p_k=1/2$ are the ones for which
the decoherence due to partition at LZ transition is the strongest.
These modes at large $t$ evolve as an equal weight superposition  
$u(t)|0\rangle + v(t)|k,-k\rangle$ with
$|u(t)|=|v(t)|$
and relative phase rapidly changing in time. 
The oscillatory phase factors will be identified below with the source
of intrinsic decoherence.

In addition, we shall see in Secs.\ref{sec:correlators1} and \ref{sec:correlators2}
that the value $z=z_\ast$, which marks the appearance of the modes
with $p_k=1/2$,
is also special in another way.
We shall find that the spin correlation functions in the final state
undergo an abrupt change at the sweep speed value corresponding to
$z=z_\ast$, from monotonic at $z<z_\ast$ to oscillatory at $z>z_\ast$.
Interestingly, this transition in the correlation function behavior
occurs at the same speed value 
which corresponds to the largest phase space of the
modes with $p_k=1/2$.

At fixed $z$, the degree of adiabaticity
for a particular mode is quite sensitive 
to the value of $k$. 
As illustrated in Fig. \ref{fig:pk}, due to the $\sin^2k$ dependence
in $p_k$,
the adiabatic regime for the modes with different $k$
is reached at different values of the sweep speed, $z\sin^2k\gg z_\ast$.
In particular, for the modes with $k$
sufficiently close to $0$ and $\pm \pi$ 
the transition is adiabatic only at very large $z$. 
These modes are special since they are 
gapless on the critical lines $h =\pm J$ 
of the equilibrium phase diagram, crossed by the evolution trajectory
(Fig. \ref{fig:phase_diagram}).
Such critical modes, characterized by small excitation frequency,
vanishing at $k=0,\pm \pi$, 
are not able to react to field sweep with finite velocity $v$, 
no matter how small the latter is. 
For the whole system, the nonadiabatic behavior of the $k=0,\pm \pi$
modes means that the spin reversal is incomplete even at very slow sweep.
The fraction of the spins that do not accomplish reversal, 
at large $z$ can be estimated as
\be\label{eq:Delta n}
\Delta n=\sum_{k\sim 0,\pm\pi}p_k\approx \frac1{\pi}\int e^{-2\pi zk^2}dk=(2\pi^2 z)^{-1/2} 
.
\ee
The density of defects $\Delta n$ has an inverse square root
dependence on the sweep speed $v$. 
By order of magnitude, the estimate (\ref{eq:Delta n}) 
can be obtained also from the momentum
value $k\simeq(z_\ast/z)^{1/2}$
corresponding to the crossover at $p_k\simeq 1/2$.

Our result (\ref{eq:Delta n}) for $\Delta n$ can be compared to the 
estimate following from the KZ causality argument \cite{kibble,zurek2},
which predicts the domains of the ordered phase
of size
\be\label{eq:L=ct}
\ell = c\tau
\ee
where $c$ is the velocity of gapless 
excitations at the critical point and $\tau$ is the 
characteristic transit time. In our case, from the excitation spectrum 
(\ref{eq:epsilon(k)}),
at the critical points $h=\pm J$ the velocity 
is $c=\gamma J$. 
The transit time for the $k$-mode
can be estimated as the time of sweeping across the gap:
$\tau_k\simeq\Delta_k/v$, where $\Delta_k=ck$. 
After identifying $\ell$ with $1/k$, Eq.(\ref{eq:L=ct}) becomes
$\ell = c^2/(v\ell)$, yielding $\ell$ vs. $v$ dependence 
\be
\label{eq:kz_length}
\ell = c/\sqrt{v}
.
\ee
The $-1/2$ power law scaling is in agreement with the result (\ref{eq:Delta n}),
which confirms the KZ scenario  \cite{kibble,zurek2}
for 1d spin chain and links it to 
the many-body LZ transition.
Similar observations were made in a recent numerical study of 
a spin model in a finite size system \cite{zurek5}.

\section{Decoherence due to Transit Through Critical Point}
\label{sec:decoherence}

Here we discuss the phenomenon of \emph{intrinsic decoherence}
resulting from massive production of spin excitations
at a sweep through critical point. 
We start with noting that 
the evolution during $-T<t<T$, taken formally,
is manifestly unitary and preserves all phase relationships. 
For the density matrix of the entire system,
the evolution $i\dot\rho=[\rho,H]$ starting with a pure state
$\rho(t=-T)=|0_N\rangle\langle 0_N|$ of $N$ spins obtains a pure state:
\be
\label{eq:full_rho}
\rho(N,T)=\rho(t=T)=\hat U_T|0_N\rangle\langle 0_N|\hat U_T^\dagger
\ee
However, we shall see that
some of the phases in the density matrix (\ref{eq:full_rho})
develop rapid oscillation at large $T$. 
The phase growing with $T$ will be found to depend on the momentum
$k$ so that the oscillatory part of $\rho$ averages
to zero after integration over $k$ in all local correlators.

It is beneficial to identify the oscillatory terms and suppress them early 
in calculation.
This can be achieved by replacing the unitarily evolved 
state (\ref{eq:full_rho}) by 
a \emph{decohered state} from which the rapidly oscillating terms
are removed. This procedure, which will be seen to describe
correctly the local properties of the evolved system,
leads to the notion of non-equilibrium 
steady state (NESS). 
Besides the benefit of simplicity, early appearance of NESS in the analysis 
also helps to develop intuition about how the results of evolution depend
on various parameters, such as the sweep rate and coupling strength.

Alternatively, one could proceed more formally,
carrying the oscillatory terms in $\rho$ over and then arguing
that they drop out in the limit of long time $T$
and large systems size. For that, one 
would have to include the effect of 
some auxiliary phyical decoherence
mechanism and obtain suppression of the oscillatory terms independent
of the strength of decoherence effect, no matter how weak the latter is.
Instead, we choose to build the NESS and its decohered density matrix
prior to analyzing the correlators.

We shall focus on the observables, i.e. spin correlators,
which are more physical quantities than the full many-body 
density matrix.
Let us consider correlators in position space
within a block $[1,n]$:
\be\label{eq:general correlator}
\la {\cal A}(x)...{\cal A}'(x')\ra
,\quad
1\le x,...,x'\le n
,
\ee
where ${\cal A}...{\cal A}'$ are local observables
given by products of a finite number of fermion operators.
In the discussion below the intermediate
length scale $n$ will be much smaller than the system size, $n\ll N$.
These correlators can be evaluated with the help of  
a reduced density matrix obtained by tracing out all spin variables outside 
the block $1\le x\le n$. The resulting density matrix
describes only the $2^n$ spin states in the block:
\be
\label{eq:reduced_rho}
\rho(n,T)=\Tr_{N-n}[\rho(N,T)]
\ee
where $\Tr_{N-n}$ 
denotes integration of the $N-n$ spins outside 
the block $1\le x\le n$,
and $T$ is evolution time. The reduced density matrix 
adequately describes the correlators and other properties
at distances shorter than $n$.

Next, we consider how by taking the three scales $N$, $T$, $n$ 
to infinity in proper order one arrives at the NESS.  
First, we take the thermodynamic limit $N\rightarrow\infty$ to eliminate recurrence times of order of level spacing for finite-size systems.
Second, we take the long time limit $T\rightarrow\infty$ to suppress
oscillations and arrive at a steady state.
Finally, we take the long-wavelength limit $n\rightarrow\infty$ and obtain
the decohered density matrix
\be
\label{eq:decohered_rho}
\rho_D=\lim_{n\rightarrow\infty}\lim_{T\rightarrow\infty}\lim_{N\rightarrow\infty}\rho_D(N,T,n)
\ee
which describes NESS in an infinite system. 

Not all phase relationships of the pure state density matrix,
Eq.(\ref{eq:full_rho}), survive the 
limiting procedure (\ref{eq:decohered_rho}).
We describe this process as decoherence by analogy with 
loss of phase information for density matrices of open quantum systems 
coupled explicitly to an environment \cite{nielsen,zurek4}.  
In contrast to the latter, however, the decoherence described by 
(\ref{eq:decohered_rho}) is of an intrinsic origin,
arising from the spin chain 
acting as both the system undergoing
decoherence and the environment that induces it.  
An implicit separation between the two emerges only when 
considering correlators in contrast
to the explicit separation in open quantum systems.

We shall use the fermion representation constructed above to evaluate $\rho_D$.
Formally, this restricts our theory to the correlators of the
form (\ref{eq:general correlator}) with the observables
${\cal A}...{\cal A}'$ all taken at equal times. 
In the fermion representation,
the full density matrix $\rho(N,T)$ decouples into a tensor product
over the $k$,$-k$ subspaces:
$\rho(N,T)=\bigotimes_{k>0}R_k$ with a $4\times4$ matrix $R_k$. The latter
has nonzero elements only between the states
$|0\ra$ and $|k,-k\ra$, since the amplitude
of the states $|\pm k\ra$, which is zero in
the initial state, cannot change with time [see Eq.(\ref{eq:Uk4times4})].
Thus within each $k$,$-k$ subspace the density matrix $R_k$ is effectively
$2\times2$, restricted to the subspace $|0\rangle$, $|k,-k\rangle$ 
where it is nonzero:
\be
\label{eq:rho_unitary}
\rho(N,T)=\bigotimes_{k>0}\rho_k
,\quad
\rho_k=\begin{pmatrix}
p_k&-q_k^*\\
-q_k&1-p_k
\end{pmatrix}
\ee
where $\rho_k$ is evaluated as $S_k|0\rangle\langle 0|S_k^{-1}$.
Here $p_k$ is given by Eq.(\ref{eq:LZprob}), and 
\be
q_k=r_k s_k e^{i(\phi_k+\eta_k)}
\ee
are obtained from the $S$-matrix (\ref{eq:asymptotic_smatrix}).

Now, let us consider correlation functions in the fermion representation.
Since the Hamiltonian is quadratic in this representation,
the state $\rho(N,T)$, obtained by evolution of the $t=-T$
fermion vacuum, is of a gaussian form. This allows to employ
Wick's theorem to write
any correlator as a sum of products of pair correlators. 
Thus an arbitrary local observable can be expressed in terms of
the $2\times2$ matrix of pair correlators 
\be
\label{eq:fermion_correlator1}
G(x,x',N,T)=\langle \vec{C}_x\vec{C}^\dagger_{x'}\rangle
\equiv \Tr[\rho(N,T)\vec{C}_x\vec{C}^\dagger_{x'}]
\ee
while $\langle\vec{C}_x\rangle=0$ for a gaussian fermion state. 

Using Eq.(\ref{eq:fermion_correlator1}) we can
obtain the decohered matrix $\rho_D$ by demanding that it reproduces 
$G(x,x',N,T)$ under the limits in Eq.(\ref{eq:decohered_rho}).
Taking $N\rightarrow\infty$ first, we write the result as an integral
over a continuous $k$ variable:
\be
\label{eq:fermion_correlator2}
G(x,x',\infty,T)=\int_{-\pi}^{\pi}\frac{dk}{2\pi}e^{-ik(x-x')}
\begin{pmatrix}
p_k&q_k\\
q_k^*&1-p_k
\end{pmatrix}
\ee
Turning to the $T\to\infty$ limit, we note that, while $p_k$
and the 
modulus $|q_k|$ 
approach the asymptotic LZ values exponentially quickly,
the phase of $q_k$ exhibits oscillations as a function of
time $T$. To the leading order, at large $T$ we have 
\be\label{eq:phaseTcosk}
\phi_k+\eta_k\approx vT^2+2JT\cos k +O(\ln T).
\ee
It is crucial that this phase has a $\cos k$ dependence on $k$.
Due to the $k$-dependent phase factor $e^{i(\phi_k+\eta_k)}$, 
with the oscillations becoming very fast at large $T$, 
the integral of $q_k$ over $k$ in Eq.(\ref{eq:fermion_correlator2})
vanishes in the limit $T\to\infty$ (which practically means $T/t_Q\gg1$). 
This argument shows that the off-diagonal elements of the correlator
$G(x,x',\infty,\infty)$ vanish for arbitrary $x$, $x'$. 
The result 
\be
\label{eq:fermion_correlator3}
G(x,x',\infty,\infty)=\int_{-\pi}^{\pi}\frac{dk}{2\pi}e^{-ik(x-x')}
\begin{pmatrix}
p_k&0\\
0&1-p_k
\end{pmatrix}
\ee
means that can simply ignore the oscillatory terms
by setting $q_k=0$ in all correlation functions. We point out that such a result
agrees with the intuition that the rapidly oscillating 
off-diagonal matrix elements of $\rho$ vanish due to arbitrarily
small decoherence, and thus can be ignored in all correlation functions.

Applying the $q_k=0$ rule to $\rho(N,T)$ given by Eq.(\ref{eq:rho_unitary})
we obtain the decohered density matrix $\rho_D$ as a product
of diagonal $2\times2$ matrices,
restricted to the subspace $|0\rangle$, $|k,-k\rangle$:
\be
\label{eq:rho_decohered}
\rho_D=\bigotimes_{k>0}\rho_{D,k}
,\quad
\rho_{D,k}=
\begin{pmatrix}
p_k&0\\
0&1-p_k
\end{pmatrix}
\ee
With such identification, the decohered pair correlator
is $G(x,x',\infty,\infty)=\Tr[\rho_D\vec{C}_x\vec{C}^\dagger_{x'}]$, as required.
The relation of the higher order correlators with the pair correlators
via Wick's theorem decomposition, including fermionic signs,
remains unchanged.

Finally, we note that $p_k=e^{-2\pi z\sin^2k}$ as a function of $k$ 
has periodicity $\pi$, while $q_k$ has periodicity $2\pi$
(see Eqs.(\ref{eq:asymptotic_r,s}),(\ref{eq:asymptotic_smatrix})). 
Thus the correlation functions of
the decohered state $\rho_D$, obtained by setting
$q_k=0$, acquire even/odd 
sublattice structure in position space.
The correlators (\ref{eq:fermion_correlator3})
vanish if $x$ and $x'$ belong to different sublattices:
\bea
\label{eq:sublattices}
\nonumber
x-x'=2n: && G(x,x')=
\begin{pmatrix}
\tilde p_n(z)&0\\
0&\delta_{n,0}-\tilde p_n(z)
\end{pmatrix}, 
\\
x-x'=2n+1: && G(x,x')=0
\eea
where
\be
\label{eq:p_n(z)}
\tilde p_n(z)=\int_{-\pi}^{\pi}p_k e^{-ikn}\frac{dk}{2\pi}
= e^{-\pi z}I_n(\pi z)
\ee
with $I_n(x)$ the modified Bessel function of the first kind.
The decoupling of the even and odd sublattice in the decohered state,
manifest in Eq.(\ref{eq:sublattices}), indicates that the decohered
density matrix factorizes as 
\be\label{eq:EOfactorization}
\rho_D=\rho_E\otimes\rho_O,
\ee
where each $\rho_{E}$, $\rho_{O}$ acts only
on the even (odd) sublattice. This factorization will be used below 
in the analysis of spin correlation functions.

\section{Entropy of the decohered state}
\label{sec:entropy_new}

The necessity of transition from the pure state 
to NESS, characterized by the decohered density matrix $\rho_D$,
can be inferred without reference to pair correlators,
by employing the procedure of coarse-graining in momentum space.
Let us consider the evolved pure state density matrix $\rho(N,T)$, 
Eq.(\ref{eq:rho_unitary}).
While the diagonal matrix elements of $\rho(N,T)$ 
are smooth functions of $k$ and
independent of $T$, 
the off-diagonal elements between $|0\rangle$ and $|k,-k\rangle$ 
rapidly oscillate as functions of both $k$ and $T$.
The oscillation $k$-dependence, described by the phase factors
$e^{\pm 2iJT\cos k}$ 
[see Eq.(\ref{eq:phaseTcosk})], becomes increasingly more rapid at 
large $T$. This property makes the oscillatory terms very sensitive to
coarse-graining in $k$ space: 
They vanish after intergrating over any  small interval
$\Delta k\ll 1$ which is large compared to $(JT)^{-1}$.
This argument, applied above to individual
correlators evaluated at finite separation
in real space using the integral representation
(\ref{eq:fermion_correlator3}), can also be applied to the entire density 
matrix.
The coarse-graining selects the matrix elements of $\rho(N,T)$
which are smooth in $k$,
suppressing the oscillating parts.
Only the diagonal elements of $\rho(N,T)$ survive in $\rho_D$,
consistent with the interpretation that
the superpositions of the $|0\rangle$ and $|k,-k\rangle$ states decohere into 
a statistical mixture.
Using the language of open quantum systems \cite{zurek4}, 
one can identify
the instantaneous eigenstates $|0\rangle$, $|k,-k\rangle$ of 
$H(t\to +\infty)$ with the pointer states which survive decoherence.

\begin{figure}
\includegraphics[width=8.5cm]{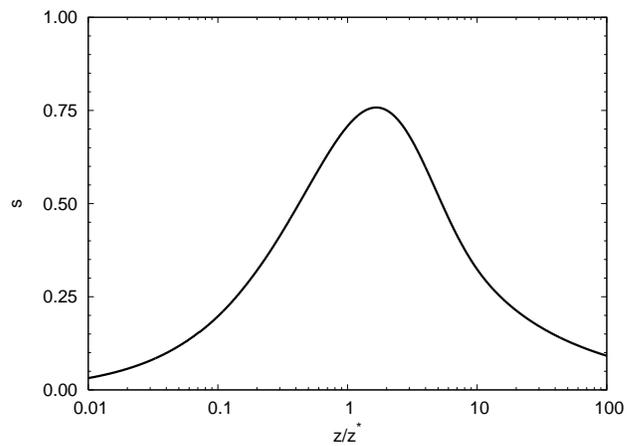}
\caption{Entropy density $s$, Eq.(\ref{eq:s_answer}), as a function of $z/z_\ast$,
inverse sweep speed.  
Note that $s$ peaks near $z_\ast$ and tends to zero for small and large $z$.}
\label{fig:entropy}
\end{figure}

To quantify the amount of information lost in the decoherence 
process \cite{zurek4},
we consider the von Neumann entropy of the system, $S=-\tr\rho_D\ln\rho_D$. 
[It will be more convenient to use natural base $\ln$ instead of a more standard
$\log_2$.]
An expression for the entropy density $s=S/N$ 
follows directly from the form (\ref{eq:rho_decohered}) of $\rho_D$:
\be
\label{eq:s}
s = -\int_{-\pi}^{\pi}\lp p_k\ln p_k +(1-p_k)\ln(1-p_k)\rp \frac{dk}{2\pi}
\ee
Using Taylor series for $\ln(1-p_k)$ and evaluating the integral 
for each term, we obtain
\be
\label{eq:s_answer}
s = (\pi z+1)\tilde p_0(z)-\pi z \tilde p_1(z)-\sum_{m=1}^{\infty}\frac{\tilde p_0(z(m+1))}{m(m+1)}
\ee
with $\tilde p_n(z)$ given by Eq.(\ref{eq:p_n(z)}).
The entropy (\ref{eq:s_answer}) as a function of sweep rate is 
plotted in Fig. \ref{fig:entropy}. We note that $s$
tends to zero in the limit of small and large $z$, since for such $z$
the dynamics gives rise to few superposition states.
The function $s(z)$ peaks near $z\simeq z_\ast$.

Let us consider the limit of slow sweep speed, $z\gg z_\ast$. In this case,
Eq.(\ref{eq:s_answer}) gives
\be\label{eq:entropy_lowv}
s= \lp \sum_{m=1}^\infty \frac1{m(m+1)^{3/2}}-\frac12\rp \Delta n
\approx 0.0761 \, \Delta n ,
\ee
where $\Delta n=(2\pi^2 z)^{-1/2}$ is the density of defects 
in the spin-reversed state (\ref{eq:Delta n}),
which describes the fraction of the spins remaining not reversed
after slow evolution. 
For fast sweeps, $z\ll z_\ast$, using the expansion $p_k=1-2\pi z\sin^2k$,
we obtain
\be\label{eq:entropy_largev}
s'=-\int_{-\pi}^{\pi} z\sin^2k \ln\lp\frac{e}{2\pi z\sin^2k}\rp
dk 
\approx \rho_0\ln\frac{c}{\rho_0}
\ee
where $\rho_0=\pi z$ is the density of the defects evaluated 
as $\int_{-\pi}^{\pi}(1-p_k)dk/2\pi$,
and $c\sim 1$ is a constant. In this case, $\rho_0$
describes the number of reversed spins, which is small at a fast sweep.

It is interesting to compare these results to the entropy of 
a classical gas of a low density $\Delta n\ll1$, 
\[
s_{\rm gas}=-\Delta n\ln\Delta n
-(1-\Delta n)\ln(1-\Delta n)\approx \Delta n\ln\frac{e}{\Delta n}.
\]
This agrees with the result 
for the fast sweep, Eq.(\ref{eq:entropy_largev}),
upon identification of $\Delta n$ with $\rho_0$.
In contrast, the value $s$ obtained for slow sweep, Eq.(\ref{eq:entropy_lowv}) 
is small compared to $s_{\rm gas}$ for the same density:
\[
s_{\rm gas}/s \approx 13.966\ln(e/\Delta n)\gg 1.
\]
Small entropy indictes that the arrangement of defects in the quantum system
after a slow sweep
is more orderly than in an ideal gas.
Another manifestation of partial ordering and correlations of the defects 
will be discussed in Secs.\ref{sec:correlators1} and \ref{sec:correlators2}, where we shall see that 
at slow sweep speeds the two-point
spin correlation function exhibits spatial oscillations,
with abrupt onset at $z=z_\ast$. As illustrated by
Fig. \ref{fig:correlator_oscill}, such oscillations
result from quasi-regular arrangement of KZ domains.

\section{\label{sec:correlators1}Spin Correlators and Toeplitz Determinants}

Here we consider the correlation functions of spin variables
$\sigma_x^\alpha$, 
and of the 
string variable $\tau_x$,
Eq.(\ref{eq:strings}), used in the fermionization
transformation.  We obtain exact expressions for these correlators
in the form of Toeplitz determinants,
which will allow to analyze them at large spatial separation.
We shall see that
the asymptotic behavior of the correlation functions
is sensitive to the sweep speed, changing abruptly
from a pure exponential decay at $z<z_\ast$
to an oscillatory dependence at $z>z_\ast$.
In this section, we focus on the non-trivial behavior for the correlators
of transverse spin $\sigma_x^1$, $\sigma_x^2$, and of $\tau_x$, 
and give a simple mathematical and physical picture. 
The detailed derivation and additional
results on $\sigma_x^3$ correlators can be found
in Sec. \ref{sec:correlators2}.

It is convenient to write the quantities of interest 
as products of Majorana fermion operators 
$A_x=a_x^\dagger+a_x$, $B_x=a^\dagger_x-a_x$.
(For convenience, we omit the factors $1/\sqrt{2}$,
$i/\sqrt{2}$
often appearing in the definition of these operators.)
The Majorana operators $A_x$, $B_x$ satisfy the algebra
\begin{eqnarray}
\label{eq:Majorana}
&& A_x^\dagger=A_x, \quad
[A_x,A_{y}]_+=\delta_{xy}\\
&& B_x^\dagger=-B_x, \quad
[B_x,B_{y}]_+=-\delta_{xy}\\
&& [A_x,B_{y}]_+=0
\end{eqnarray}
In the fermion representation, the pair products of the spin variables
$\sigma_x^\alpha$ as well as the string variables $\tau_x$,
appearing in the correlators, 
can be expressed as products of Majorana operators as follows:
\begin{eqnarray}
\label{eq:sigma_Majorana1}
\sigma^1_x\sigma^1_{x+n}&=&B_{x}A_{x+1}B_{x+1}\ldots A_{x+n-1}B_{x+n-1}A_{x+n}\\
\label{eq:sigma_Majorana2}
\sigma^2_x\sigma^2_{x+n}&=&A_{x}A_{x+1}B_{x+1}\ldots A_{x+n-1}B_{x+n-1}B_{x+n}\\
\label{eq:sigma_Majorana3}
\tau_x\tau_{x+n}&=&A_{x}B_{x}A_{x+1}B_{x+1}\ldots A_{x+n}B_{x+n}
.
\end{eqnarray}
To obtain expectation values,
we average the products of a finite number of the operators $A_x$, $B_x$
using Wick's theorem and 
the decohered density matrix $\rho_D$, Eq.(\ref{eq:rho_decohered}),
introduced in Sec.\ref{sec:decoherence}.

An additional simplification occurs due to decoupling
of the fermionic correlators, 
evaluated with the decohered density matrix $\rho_D$,
into a product of separate contributions
of the even and odd sublattice, Eq.(\ref{eq:EOfactorization}).
Let us explore this factorization 
for the correlator $\la \sigma_x^1\sigma_{x+2n}^1\ra$. 
By regrouping the operators $A_x$, $B_x$, 
separating the parts corresponding to the two sublattices,
we write 
\begin{eqnarray}
\nonumber
\sigma^1_x\sigma^1_{x+2n}&=&(B_{x}A_{x+2}B_{x+2}\ldots A_{x+2n-2}B_{x+2n-2}B_{x+2n})\\
\label{eq:factorization1}
&&(A_{x+1}B_{x+1}\ldots A_{x+2n-1}B_{x+2n-1}).
\end{eqnarray}
Comparing the two expressions in parentheses to Eqs.(\ref{eq:sigma_Majorana2}),(\ref{eq:sigma_Majorana3}),
we see that the spin operator pair
product $\sigma_x^1\sigma_{x+2n}^1$ 
evaluated on the {\it full lattice} is a product of
analogous operators $\sigma_x^1\sigma_{x+n}^1$ and $\tau_x\tau_{x+n}$,
each evaluated on a {\it sublattice}.
This leads to factorization for the expectation values since 
fermionic pair correlators do not mix different 
sublattices. The result can be symbolically written as
\be
\label{eq:corr_factorization1}
\langle\sigma^1_x\sigma^1_{x+2n}\rangle=
\langle\langle\sigma^1_x\sigma^1_{x+n}\rangle\rangle
\langle\langle\tau_x\tau_{x+n}\rangle\rangle
\ee
where the brackets $\la...\ra$ 
describe expectation values of operators on the full lattice,
while $\la\la...\ra\ra$ refer to an expectation value on a sublattice.
Similar reasoning for other correlators leads to 
\begin{eqnarray}
\label{eq:corr_factorization2}
\langle\sigma_x^2\sigma_{x+2n}^2\rangle&=&
\langle\langle\sigma_x^2\sigma_{x+n}^2\rangle\rangle
\langle\langle\tau_x\tau_{x+n}\rangle\rangle\\\
\label{eq:corr_factorization3}
\langle\tau_x\tau_{x+2n}\rangle&=&
\langle\langle\tau_x\tau_{x+n}\rangle\rangle
\langle\langle\tau_x\tau_{x+n}\rangle\rangle
\end{eqnarray}
where single (double) brackets refer to correlators
on the full lattice (sublattice).  This allows us to focus
just on the sublattice correlators.

With the help of fermionization, the sublattice 
correlators at separation $n$ can be written 
in terms of $n\times n$ 
determinants of Toeplitz matrices, defined
by a set of constant diagonals:
\be
\label{eq:toeplitz_det1}
D_n[f]=\det
\begin{vmatrix}
f_0   &f_{-1}&\cdots&f_{-(n-1)}\\
f_1   &f_0   &\cdots&f_{-(n-2)}\\
\vdots&\vdots&\ddots&\vdots    \\
f_{n-1}&f_{n-2}&\cdots&f_0
\end{vmatrix}
\ee
The structure of the matrix, completely specified by the set of 
numbers $f_n$, can be encoded in a generating function
\be
\label{eq:gen_function}
f(\xi)=\sum_n f_n\xi^n,
\quad f_n=\oint_C\frac{d\xi}{\xi}\xi^{-n}f(\xi)
\ee
with the contour $C$ being the unit circle $|\xi|=1$.
The properties of Toeplitz determinants depend on the 
combination of poles, zeros and other singularities of $f(\xi)$
in the complex plane \cite{bottcher1}.

In our case, the Toeplitz matrix representation is obtained
by evaluating the sublattice correlators in (\ref{eq:corr_factorization1}),
(\ref{eq:corr_factorization2}), (\ref{eq:corr_factorization3})
using fermion representation. With the help of Wick's theorem,
all correlators can be expressed as polynomials of pair correlators
of Majorana fermions. Due to the sublattice structure of $\rho_D$, the nonzero
pair averages are all of the form $\la a_x^\dagger a_{x'}\ra$ with $x-x'$ even.
In addition, the expectation values $\la A_xA_{x'}\ra$ and
$\la B_xB_{x'}\ra$, with $x\ne x'$, are zero due to Majorana fermion algebra. 
Only the pairs of operators $B_xA_{x'}$ 
give nonzero expectation values:
\be
\la B_xA_{x'}\ra =\int_{-\pi}^{\pi} e^{ik(x-x')}(1-2p_k)\frac{dk}{2\pi}
\ee 
where $p_k$ is the LZ probability, Eq.(\ref{eq:LZprob}).
[We note that, while $A_xA_{x'}=B_xB_{x'}=1$ at $x=x'$,
such combinations do not arise in the fermionic representation 
of spin variables.]
Summing over all pair contractions with appropriate fermionic signs 
brings the sublattice spin
correlators to the Toeplitz determinant form:
\begin{eqnarray}\label{eq:correlators_Dn1}
\langle\langle\sigma^1_x\sigma^1_{x+n}\rangle\rangle&=&D_n[g^{+1,z}]\\
\label{eq:correlators_Dn2}
\langle\langle\sigma^2_x\sigma^2_{x+n}\rangle\rangle&=&D_n[g^{-1,z}]\\
\label{eq:correlators_Dn3}
\langle\langle\tau_x\tau_{x+n}\rangle\rangle&=&D_n[g^{0,z}]
\end{eqnarray}
where the generating functions $g^{m,z}$ are defined as
\be\label{eq:g^mz}
g^{m,z}(\xi)=-(-\xi)^m ( 1-2 p_k)
,\quad
\xi=e^{2ik}
.
\ee
This form of the generating function, and, in particular,
the origin of the factors $\xi$, $\xi^{-1}$, can be understood as follows.
The string of $A_xB_x$ operators appearing in 
the $\sigma_x^1$ correlator has an additional $B_x$ at 
the beginning and $A_{x+n}$
at the end
compared to a similar string for
the $\tau$ correlator.
This results in a shift of the matrix elements $g_n\rightarrow g_{n+1}$
in the determinants for $\tau$ compared to the one for $\sigma_x^1$,
which translates to the mapping $g(\xi)\rightarrow \xi g(\xi)$
of the generating functions.  
Similar reasoning accounts for the factor $\xi^{-1}$ for 
the $\sigma_x^2$
correlator generating function.
The factor of $2$ in the relation 
$\xi=e^{2ik}$ arises because 
the correlators are restricted to a sublattice, which makes the $k$-dependence 
$\pi$ periodic rather than $2\pi$ periodic.  The factor $-(-1)^m$ ensures 
correct fermionic sign. 

The Toeplitz matrix representation allows to study the correlation 
functions numerically, since evaluating determinants on a computer
is a low cost operation. However, as we show below,
the problem can also be handled analytically. The benefit of the analytic treatment is that it provides a very clear and complete description
of the behavior of the correlation functions at different sweep speeds, 
including the transition at $z=z_\ast$.

\section{Spin correlators asymptotics}

We are primarily interested in the behavior 
of the sublattice correlators at large separation which maps
to the large-$n$ asymptotics of Toeplitz determinants. 
It is instructive to recall the Szeg\"o limit theorem
result for the Toeplitz determinant (\ref{eq:toeplitz_det1})
asymptotic behavior:
\be\label{eq:Szego_thm}
D_n[f]\approx \exp\lp n\int_0^{2\pi}\ln f(e^{i\theta})\frac{d\theta}{2\pi}\rp
\ee
which holds when the generating function 
$f(\xi)$ has a zero winding number and no singularities on the unit circle.
The origin of the asymptotic (\ref{eq:Szego_thm}) can be seen by noting that
in this case the matrix elements $f_n$ rapidly decrease with $|n|$,
and the Toeplitz matrix can be approximated by a band matrix.
Then the result (\ref{eq:Szego_thm}) naturally follows after closing the interval
$1\le x\le N$ into a circle and going to Fourier representation. 
The question of how the asymptotic (\ref{eq:Szego_thm}) is modified in 
the cases when the winding numbers are nonzero and/or the generating function
has singularities on the unit circle has been a subject of many publications.
Not trying to review all the literature, in the discussion below we will 
cite the available results, either conjectured or proven, as appropriate.

We shall start with the simplest situation, 
for which Szeg\"o limit theorem provides a suitable framework.
Let us consider the Toeplitz determinant representation
for the correlator (\ref{eq:correlators_Dn3})
with the generating function $f(\xi)=g^{0,z}(\xi)$. This function is
real for $|\xi|=1$, and thus has zero winding number.
In this case, Eq.(\ref{eq:Szego_thm}) yields 
\[
D_n[g^{0,z}]\approx
e^{an},
\quad
a=\int_0^\pi \ln\lp 1-2 e^{-2\pi z\sin^2k}\rp \frac{dk}{\pi}
\]
The expression for $a$ is analytic at $z<z_\ast$, 
has a singularity at $z=z_\ast$,
and becomes ill-defined at $z>z_\ast$.
To clarify the origin of this behavior, let us inspect zeros of $g^{0,z}$.
There is an infinite number of zeros $\xi=\lambda_p,\lambda_p^{-1}$ 
of multiplicity one, with $p$ an integer, 
which can be found from the representation
\be
g^{0,z}(\xi)=e^{-\pi z(1-2z_\ast/z-x)}-1
,
\ee
where $x=(\xi+\xi^{-1})/2$. We obtain
\be
\label{eq:roots}
\lambda_p=\exp\lb -\acosh\left(1-\frac{\ln 2}{\pi z}-\frac{2ip}{z}\right)\rb
\ee
where we choose the branch of $\acosh(x)$ with positive real part so that
$|\lambda_p|\le1$.  
Note that $|\lambda_p|>|\lambda_p'|$ for $|p|<|p'|$, so that the zeros closest
to the unit circle are $\lambda_0$ and $\lambda_0^{-1}$,
which satisfy
\be\label{eq:lambda0}
\frac12 (\lambda+\lambda^{-1})=1-2 z_\ast/z
.
\ee
The $\lambda(z)$ dependence has a square root singularity at $z=z_\ast$.
To specify the analyticity branch near the singularity, we take
$p=+0$, with an infinitesimal positive
part, in Eq.(\ref{eq:roots}).

\begin{figure}
\begin{center}
\psfrag{l}{$\lambda_0$}
\psfrag{l1}{$\lambda_0^{-1}$}
\psfrag{z=zs}{$z=z_\ast$}
\includegraphics[width=2.5in]{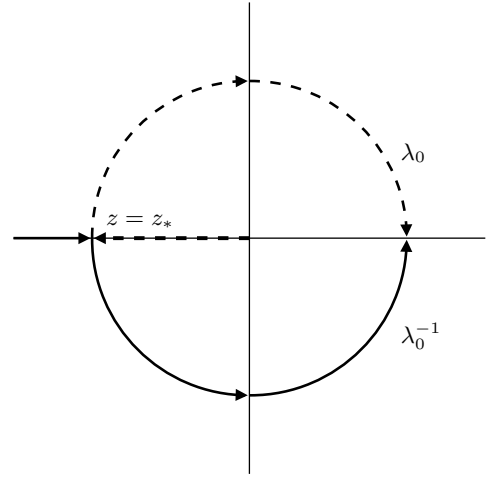}
\caption{ \label{fig:roots_motion}
Motion of the roots $\lambda_0$, $\lambda_0^{-1}$ as a function of $z$ from the
negative real axis for $z<z_\ast$ to the unit circle for $z>z_\ast$.  The 
 direction of the arrows indicate increasing $z$.
}
\end{center}
\end{figure}

The $z$ dependence of the roots (\ref{eq:lambda0})
is illustrated in Fig.\ref{fig:roots_motion}.
Both roots are real and negative
at $z<z_\ast$: $\lambda_0<-1<\lambda_0^{-1}<0$. As $z$ tends to $z_\ast$,
the roots move along the real axis towards $\xi=-1$,
approach one another and merge
at $z=z_\ast$, then split and remain on the unit circle at $z>z_\ast$,
with $\lambda_0^{-1}=\lambda_0^\ast$. 
This leads to a singularity of the determinants $D_n[g^{0,z}]$,
and thus of the correlation functions, at $z=z_\ast$.

To better understand the behavior near $z=z_\ast$, it is instructive 
to try isolate the effect of the roots $\lambda_0$, $\lambda_0^{-1}$.
For that, we consider simplified generating functions
\be
\label{eq:simple_f}
f^{(m)}(\xi)=
-(-\xi)^m\lambda_0^{-1}
\left(1-\lambda_0\xi\right)
\left(1-\lambda_0\xi^{-1}\right)
\ee
where $m=0,\pm 1$ and $\lambda_0$, $\lambda_0^{-1}$ 
are defined by Eq.(\ref{eq:lambda0}). 
The simplified expressions (\ref{eq:simple_f}) 
capture most of the non-trivial behavior 
of the sublattice correlators arising at $z\approx z_\ast$.
Each of the functions $f^{(m)}$ has only three nonzero
Fourier coefficients $f^{(m)}_n$, 
and thus the Toeplitz matrix in this case is three-diagonal.
One can easily calculate the corresponding Toeplitz determinants, obtaining
\begin{eqnarray}
\label{eq:simple_det1}
&& D_n[f^{(\pm1)}] = (-1)^n\\
\label{eq:simple_det2}
&& D_n[f^{(0)}] =
(-1)^n\frac{\lambda_0^{n+1}-\lambda_0^{-(n+1)}}{\lambda_0-\lambda_0^{-1}}.
\end{eqnarray}
These quantities, obtained from the simplified generating
functions, Eq.(\ref{eq:simple_f}), 
describe the qualitative behavior of the sublattice
correlators for $\sigma^1_x$, $\sigma^2_x$, and $\tau_x$, 
according to 
Eqs.(\ref{eq:correlators_Dn1}),(\ref{eq:correlators_Dn2}),(\ref{eq:correlators_Dn3}).

The expressions (\ref{eq:simple_det1}) are independent of $\lambda_0$,
indicating a smooth behavior of
$\sigma^1_x$, $\sigma^2_x$ sublattice correlators 
with $z$ which will persist upon including the full generating function.
The $m=0$ determinant, Eq.(\ref{eq:simple_det2}), 
is analytic as a function of $\lambda_0$ even at $\lambda_0=\lambda_0^{-1}$.
More interestingly, and somewhat unexpectedly, 
it is analytic as a function of $z$ at $z=z_\ast$, since the right hand side
of Eq.(\ref{eq:simple_det2}) is polynomial in 
$\lambda_0+\lambda_0^{-1}$.
As a function of $n$, the expression (\ref{eq:simple_det2})
exhibits a crossover from exponential behavior at $z<z_\ast$
to oscillatory behavior as $z>z_\ast$.
In addition,
it grows linearly with $n$ exactly at $z=z_\ast$. 
This crossover behavior, as well as nonanaliticity in $z$, 
persist upon including the full generating function.  

For comparison, let us consider the asymptotics for sublattice correlators 
obtained from the full generating function,
as discussed in Sec. \ref{sec:correlators2}:
\begin{eqnarray}
\label{eq:sub_corr1}
&& \langle\langle\sigma_x^\alpha\sigma_{x+n}^\alpha\rangle\rangle \approx E_1 (-G)^n
,\quad (\alpha=1,2)
\\
\label{eq:sub_corr2}
&& \langle\langle\tau_x\tau_{x+n}\rangle\rangle \approx
E_1(-G)^n\frac{\lambda_0^{n+1}E_2-\lambda_0^{-n-1}E_2^{-1}}{\lambda_0-\lambda_0^{-1}}
\end{eqnarray}
where $G$ and $E_{1,2}$,
given by Eqs.(\ref{eq:corr_paramG}),(\ref{eq:corr_paramE1}),(\ref{eq:corr_paramE2}), 
have a smooth $z$ dependence.  
We note similarity of the behavior of
these expressions to Eqs.(\ref{eq:simple_det1}),(\ref{eq:simple_det2})
at $\lambda_0\approx\lambda_0^{-1}$. 
We see that the origin
of the crossover behavior in the sublattice correlators, 
resulting in nonanaliticity in $z$, is indeed the motion of
the zeros $\lambda_0$ from the real axis to the unit circle.

\begin{figure}
\begin{tabular}[b]{cc}
\includegraphics[width=3.0in]{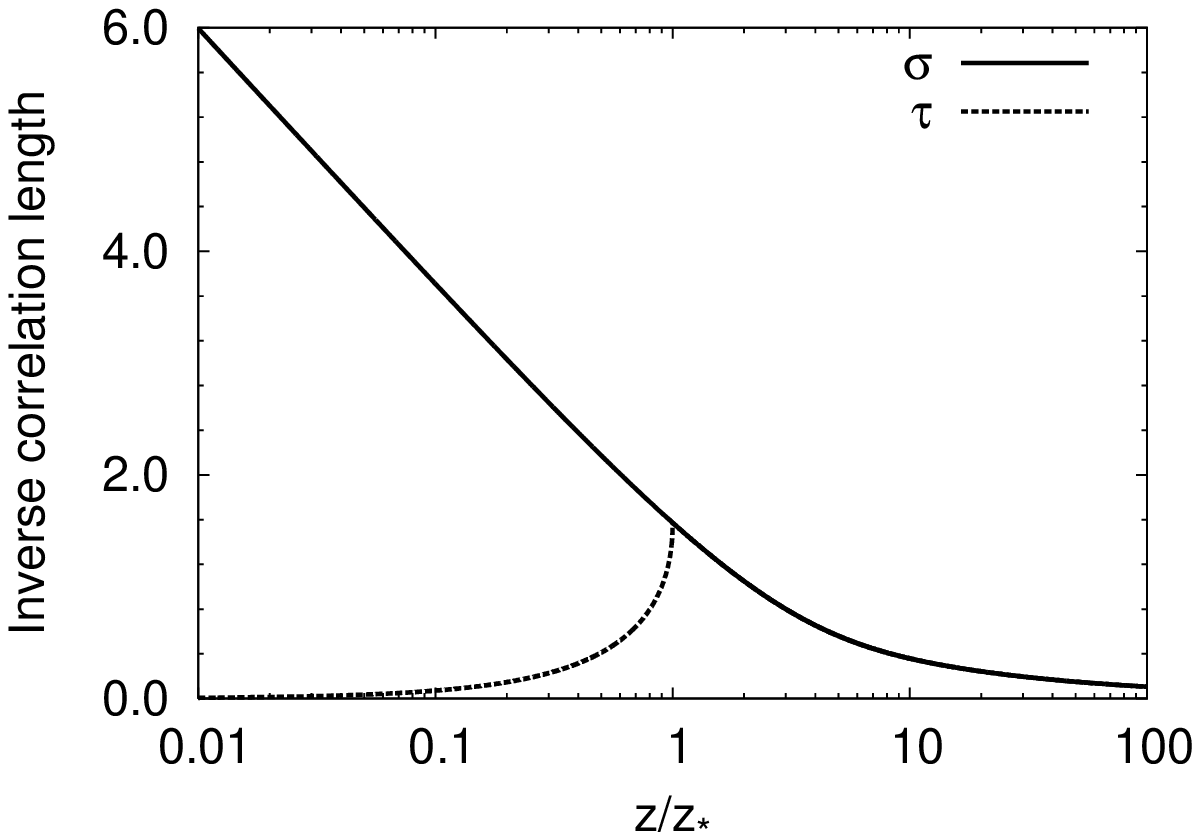}&\textbf{(a)}\\
\includegraphics[width=3.0in]{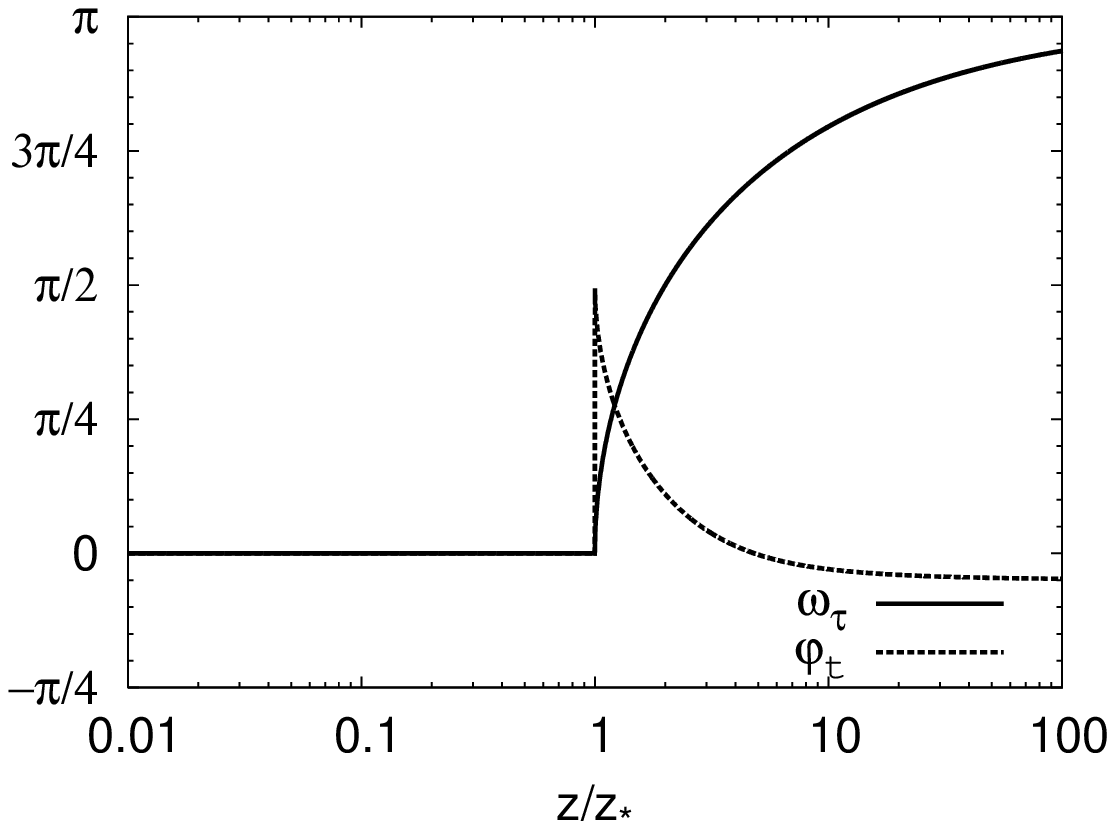}&\textbf{(b)}\\
\includegraphics[width=3.0in]{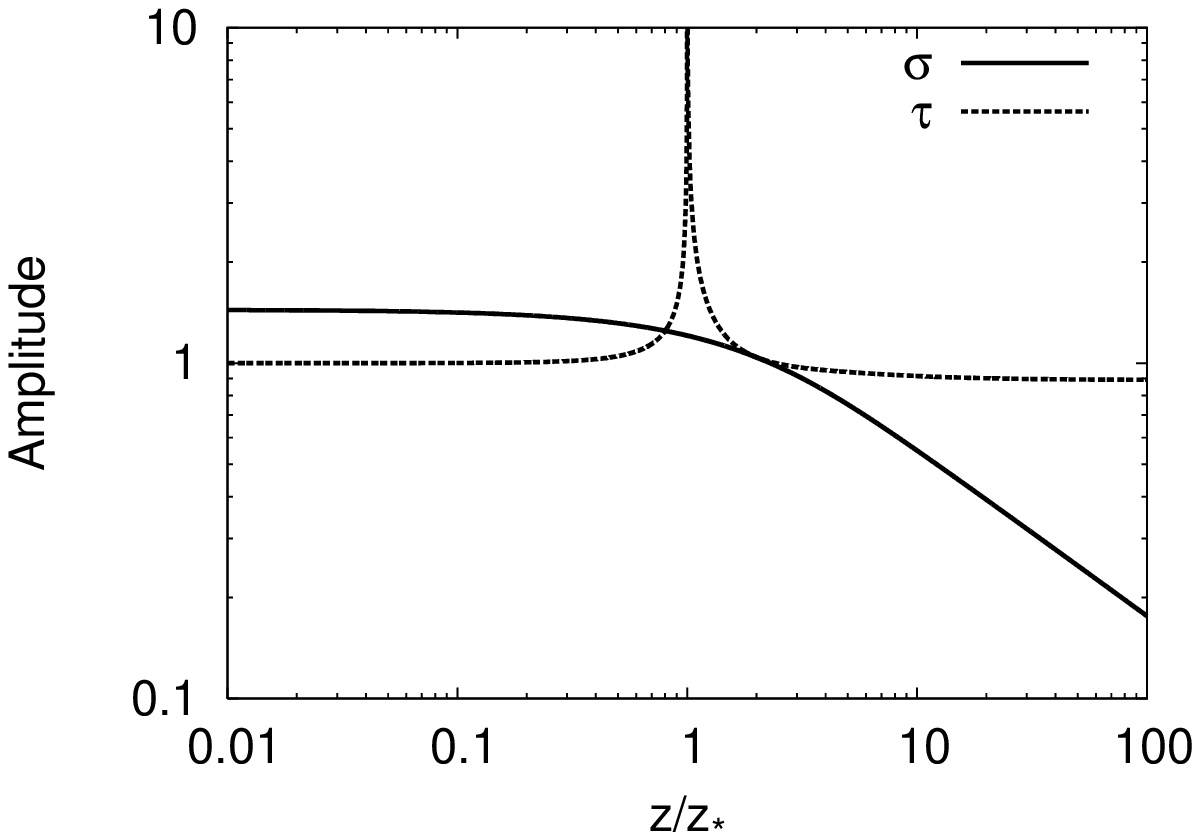}&\textbf{(c)}\\
\end{tabular}
\caption{The sublattice correlation function
parameters, Eqs.(\ref{eq:sub_corr_canonical1}),(\ref{eq:sub_corr_canonical2}),
dependence on the inverse sweep speed $z/z_\ast$: 
(a) the correlation
lengths $\ell_\sigma$, $\ell_\tau$; 
(b) the frequency $\omega_\tau$ and phase shift $\phi_\tau$;
(c) the amplitudes $A_\sigma$, $A_\tau$.
Shown are the analytical dependences obtained from 
Eqs.(\ref{eq:sub_corr1}),(\ref{eq:sub_corr2}), which were verified
by evaluating Toeplitz determinants numerically.
}
\label{fig:correlator_plot}
\end{figure}

It will be useful to also write the sublattice correlators in the canonical form
\begin{eqnarray}
\label{eq:sub_corr_canonical1}
&& \langle\langle\sigma_x^{\alpha}\sigma_{x+n}^{\alpha}\rangle\rangle \approx 
A_{\sigma}
e^{-n/\ell_\sigma}
\cos \pi n \\
\label{eq:sub_corr_canonical2}
&& \langle\langle\tau_x\tau_{x+n}\rangle\rangle \approx A_\tau
e^{-n/\ell_\tau}
\cos(\omega_\tau n-\phi_\tau)
\end{eqnarray}
($\alpha=1,2$). The parameters appearing in these expressions,
the amplitudes $A_{\sigma,\tau}$, 
the correlation lengths $\ell_{\sigma,\tau}$, 
the wavenumber of spatial oscillations 
$\omega_\tau$, and the phase $\phi_\tau$, are plotted  
as a function of $z/z_\ast$ in Fig. \ref{fig:correlator_plot}.

The correlation lengths 
$\ell_\sigma$ and $\ell_\tau$ both become large at a slow sweep speed.
At a fast sweep, the $\sigma^{1,2}_x$ correlators become short-ranged,
while the $\tau_x$ correlator is long-ranged.
The oscillatory behavior of the $\tau_x$ correlator appears abruptly
at $z=z_\ast$, with the spatial frequency and other parameters 
displayed in Fig. \ref{fig:correlator_plot} having non-analytic behavior.
The character of this singularity is similar to that exhibited by the simplified
model discussed above, Eqs.(\ref{eq:simple_det1}),(\ref{eq:simple_det2}),
which is controlled by the zeroes of the generating function
nearest to the unit circle.

Although the sublattice correlators are mathematically convenient, 
the physical content of our results
becomes more transparent in the full lattice correlators.
From the factorization relation, Eq.(\ref{eq:corr_factorization1}), 
since $\cos \pi n=(-)^n$, the $\sigma^{1,2}$ correlators are simply given by
\be
\label{eq:full_corr_canonical}
\langle\sigma_x^{\alpha}\sigma_{x+2n}^{\alpha}\rangle\approx A_{\sigma}A_\tau
e^{-n/\ell}\cos((\pi-\omega_\tau)n+\phi_\tau)
\ee
$(\alpha=1,2)$, where $\ell^{-1}=\ell_{\sigma}^{-1}+\ell_{\tau}^{-1}$.  

Now, let us discuss the physical regimes described by these correlations 
functions. In the time evolution considered here, the system is driven
from the disordered through the ordered phase and back into the disordered phase.
In equilibrium, the correlations of $\sigma_x^{1,2}$ 
are absent at the early and late times,
i.e. in the disordered phases,  
but can build up at intermediate times when the system is in the ordered phase.
The simplest situation arises at small $z$, i.e. high sweep speed.
In this case, 
all the modes in the system can be treated in a sudden approximation. 
There is very little time for correlations in the ordered phase to build up, 
which results in very short range correlations
described by exponential decay with a small correlation length.

In contrast, large $z$  describes the slow sweep speed regime,
when the dynamics becomes more adiabatic. 
However, full adiabaticity cannot be reached
for a system driven across quantum critical points
where the gap vanishes.  
The build-up of correlations upon crossing the first quantum critical point 
from the disordered to ordered phase, $h=-J$, 
can be understood in the KZ framework as appearance of ordered domains 
of size $\ell\approx c/\sqrt{v}$, Eq.(\ref{eq:kz_length}).
The length scale $\ell$ characterizes separation between defects 
of the ordered state, resulting from nonadiabaticity
at crossing the critical point.  
The defects for the ferromagnetically ordered state, 
describing our system in equilibrium at $-J<h(t)<J$, 
are domain walls separating domains with opposite magnetization.
The magnetization sign alternation in the domains
leads to an oscillatory behavior of the correlators
on top of exponential decay, as illustrated in Fig. \ref{fig:correlator_oscill}.  
Subsequent crossing into the disordered phase at $h=J$
then leads to suppression 
of the correlations built up in the ordered phase.  
This is consistent with the
behavior of the full lattice correlators where both the 
correlation length $\ell$ and the spatial period $2\pi/(\pi-\omega_\tau)$
grow as $z^{1/2}$ at large $z$, 
while the correlator amplitude $A_\tau A_\sigma$ goes to zero.
 
The crossover between the two regimes, occurring at $z=z_\ast$, 
corresponds to sweep speed of the order
of the inverse bandwidth.  
As in the discussion of Eqs.(\ref{eq:simple_det1}),(\ref{eq:simple_det2}), 
the behavior of the correlation functions
near $z=z_\ast$ at fixed $n$ is analytic and is described
as a smooth crossover.  
The apparently singular behavior in Fig. \ref{fig:correlator_plot} 
is analogous to
Stoke's phenomenon \cite{erdelyi} for asymptotic series,
where the coefficients of an asymptotic expansion of a function
may not be analytic in some parameters even when the function itself 
is analytic in those parameters.

\section{\label{sec:correlators2}Spin Correlators II}

In this section we outline the details of derivation of 
the results discussed above.
A general procedure for calculating the asymptotics for Toeplitz determinants 
from the structure of the singularities in the generating function 
is described in Appendix \ref{app:toeplitz}.
We note that, while this procedure in its most general form is 
only a conjecture, it is a 
reasonable extension of known rigorous results.  
Moreover, since our generating function, Eq.(\ref{eq:g^mz}), 
has only simple zeros of integer order, 
the approach used below stands on a firm ground: In this case,
as discussed in Appendix \ref{app:toeplitz}, 
our procedure 
follows from a rigorous result of Ref.\cite{bottcher2}.
In addition, we have compared our analytic results to
the correlation functions obtained from direct numerical evaluation of 
the Toeplitz determinants, and found them to be in full agreement.

As discussed above, among all roots of our generating function,
Eq.(\ref{eq:roots}), one pair, $\lambda_0$ and $\lambda_0^{-1}$, plays a special role.
We write the generating function $g^{m,z}$ in the form
\be
\label{eq:full_gen_function2}
g^{m,z}(\xi)=
-(-\xi)^m\lambda_0^{-1}
\left(1-\lambda_0\xi\right)
\left(1-\lambda_0\xi^{-1}\right)
e^{h(\xi)}
\ee
which isolates these most relevant roots
into a factor identical to the simplified generating function
discussed above, Eq.(\ref{eq:simple_f}). 
The remaining part, $e^{h(\xi)}$, has the form
\be
\label{eq:smooth_part}
e^{h(\xi)}=e^{\pi z\left(\xi+\xi^{-1}\right)/4}\frac{\pi z}{\sqrt{2e^{\pi z}}}
\prod_{p\ne0}\frac{z(1-\lambda_p\xi)(1-\lambda_p\xi^{-1})}{4|p|\lambda_p}
\ee
It is explicit in this expression that $e^{h(\xi)}$ 
has all its singularities located 
further away from the unit circle than 
$\lambda_0$ and $\lambda_0^{-1}$. The expression for $h(\xi)$
can be written in a more compact form:
\be
%% \tilde{h}(x)
h(\xi)=
\ln\lp
\frac{1-e^{-\pi z(1-2z_\ast/z-x)}}{2(1-2z_\ast/z-x)}
\rp
\ee
where $x=(\xi+\xi^{-1})/2$.

We obtain the correlator asymptotics given in
Eqs.(\ref{eq:sub_corr1}),(\ref{eq:sub_corr2}) either by using the result of 
Ref. \cite{bottcher2} or by the more general method of Appendix \ref{app:toeplitz}.
The latter procedure involves a contour $C$ that
passes through the two roots $\lambda_0$, $\lambda_0^{-1}$ closest to the unit circle.
(This contour does not have to be a circle when $|\lambda_0|\ne 1$.)
We isolate the contributions of $\lambda_0$ and $\lambda_0^{-1}$,
and incorporate the rest of the generating function into a part smooth off of $C$, 
denoted by $h(\xi)$. In the contribution
of $h(\xi)$ to the quantities in the asymptotics,
given by contour integrals over $C$,
we can deform $C$ to the unit circle. Finally,
we reparameterize
the complex variable $\xi$ on the unit circle
in the contour integral with $x=\cos\theta$, $\xi=e^{i\theta}$.
This yields expressions for the 
parameters $G$, $E_i$ of the form 
\begin{widetext}
\begin{eqnarray}
\label{eq:corr_paramG}
\ln G&=&h_0=\int_{-1}^{1}\frac{dx}{\sqrt{1-x^2}}\tilde{h}(x)\\
\label{eq:corr_paramE1}
\ln E_1&=&\sum_{n=1}^{\infty}nh_n^2=\ln\left(\frac{2G}{\pi z}\right)+
\frac{1}{2\pi^2}\dashint_{-1}^{1}\dashint_{-1}^{1}dxdy\tilde{h}'(x)
\frac{\tilde{h}(x)-\tilde{h}(y)}{x-y}
\sqrt{\frac{1-x^2}{1-y^2}}\\
\label{eq:corr_paramE2}
\ln E_2&=&\sum_{n=1}^{\infty}h_n(\lambda_0^n-\lambda_0^{-n})=-\ln\left(\frac{\pi z^3G}{32}\right)\Theta(z_\ast-z)
-\frac{1}{\pi}\dashint_{-1}^{1}\frac{dx}{\sqrt{1-x^2}}\tilde{h}(x)\frac{(\lambda_0-\lambda_0^{-1})/2}{(\lambda_0+\lambda_0^{-1})/2-x}
\end{eqnarray}
\end{widetext}
where and $\Theta(x)$ is the step function.
Here $h_n$ are the Fourier coefficients of $h(\xi)=\sum_n h_n\xi^n$, and the function $\tilde h(x)$
%% \be
%% \tilde{h}(x)=\ln\left[\frac{1-e^{-\pi z(1-2z_\ast/z-x)}}{2(1-2z_\ast/z-x)}\right]
%% \ee
%% %
is just $h(\xi)$ reparameterized with $x=(\xi+\xi^{-1})/2$.  
The integrals are in the principal value sense where appropriate.
Both $G$ and $E_1$ are positive and smooth in $z$.  However, $E_2$ is real for $z<z_\ast$ and 
a pure phase for $z>z_\ast$ which is the same behavior as $\lambda_0$.

In passing from correlators in the form of Eqs.(\ref{eq:sub_corr1}),(\ref{eq:sub_corr2})
to the canonical form of 
Eqs.(\ref{eq:sub_corr_canonical1}),(\ref{eq:sub_corr_canonical2})
we have to formally drop the $\lambda_0^n$ for $z<z_\ast$ since it is subleading compared to $\lambda_0^{-n}$.
This gives the following relation between the parameters:
\begin{eqnarray}
\label{eq:canonical_param1}
A_\sigma&=&E_1
\\
\label{eq:canonical_param2}
\ell^{-1}_\sigma&=&-\ln G\\
\label{eq:canonical_param3}
\ln A_\tau&=&\Re\ln\left(\frac{E_1}{E_2(1-\lambda_0^2)}\right)\\
\label{eq:canonical_param4}
\ell^{-1}_\tau&=&-\Re\ln\left(\frac{G}{\lambda_0}\right)\\
\label{eq:canonical_param5}
\omega_\tau&=&(\pi-\Im\ln\lambda_0)\Theta(z-z_\ast)\\
\label{eq:canonical_param6}
\phi_\tau&=&(\Im\ln\left(\lambda_0 E_2\right)-\pi/2)\Theta(z-z_\ast)
\end{eqnarray}
where $\alpha=1,2$ and $\Theta(x)$ is the step function.
We note that $\omega_\tau$ and $\phi_\tau$ are nonzero only at $z>z_\ast$.
Also, the two correlation lengths $\ell_\sigma$ and $\ell_\tau$ are equal 
at $z>z_\ast$ and differ at $z<z_\ast$ (see Fig.\ref{fig:correlator_plot}).

Now, let us consider the behavior at large $z$,
describing slow sweep speed.
As we noted earlier, in this case the length scales 
$\xi_\sigma$, $\xi_\tau$ and $2\pi/(\pi-\omega_\tau)^{-1}$ 
are comparable to
the typical length scale separating KZ defects (i.e. domain walls).
These quantities are given by $\ln G$ and $\ln \lambda_0$
via Eqs.(\ref{eq:canonical_param2}),(\ref{eq:canonical_param4}),(\ref{eq:canonical_param5}).
Eq.(\ref{eq:roots}) gives $\ln \lambda=-\acosh(1-2z_\ast/z)$ which has a known large-$z$ expansion.
Using the integral representation in 
Eqs.(\ref{eq:corr_paramG}),(\ref{eq:corr_paramE1}),(\ref{eq:corr_paramE2}) we obtain
\be
\ln G=\ln(1-2e^{-\pi z})-\frac{2}{\pi}\dashint_{0}^{2\pi z}dx\frac{\acos(1-x/\pi z)}{e^x-2}
\ee
after integrating once by parts and changing variables $x\rightarrow\pi z(1-x)$.  For large $z$,
we can drop the $\ln$ term, which is exponentially small,
expand $\acos$ in $1/z$, and integrate term by term with 
the upper limit at infinity to obtain the large-$z$ expansion.
This procedure yields
\begin{eqnarray}
&& \ell^{-1}_\sigma =\ell^{-1}_\tau
\approx
%%\frac{1}{\sqrt{2\pi^4 z}}\sum_{n=0}^{\infty}\frac{\Gamma(n+1/2)^2\Re\Li_{n+3/2}(2)}{\Gamma(n+1)}
\sum_{n=0}^{\infty}\frac{A_n}{(2\pi z)^{n+1/2}}\\
&&\omega_\tau \approx \pi-
%% \sqrt{\frac{\ln 2}{2\pi^2 z}}\sum_{n=0}^{\infty}\frac{\Gamma(n+1/2)}{(n+1/2)\Gamma(n+1)}
\sum_{n=0}^{\infty}B_n\left(\frac{\ln 2}{2\pi z}\right)^{n+1/2}
.
\end{eqnarray}
Here the coefficients $A_n$, $B_n$ are given by
\bea
&& A_n=\frac{\Gamma(n+\frac12)^2\Re\Li_{n+3/2}(2)}{\pi^{3/2}\Gamma(n+1)}\\
&& B_n=\frac{\Gamma(n+\frac12)}{\pi^{1/2}(n+\frac12)\Gamma(n+1)}
\eea
where $\Li_\nu(x)$ is the polylogarithm function and 
$\Gamma(x)$ is the gamma function.
These expressions exhibit the scaling $\ell_{\sigma,\tau},\,(\pi-\omega)^{-1}\propto z^{1/2}$ 
expected from the KZ picture.

\section{Magnetization}
\label{sec:magnetization}

\begin{figure}
\includegraphics[width=8.5cm]{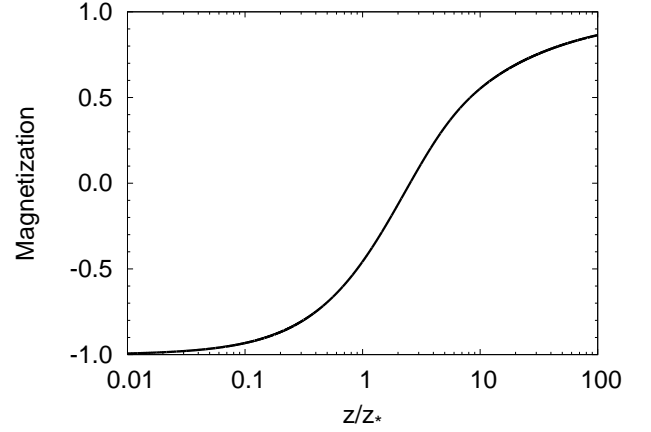}
\caption{
Magnetization $m^z$, Eq.(\ref{eq:m^z}), as a function of 
the inverse sweep speed $z/z_\ast$.
Note the smooth transition from small to large $z$.}
\label{fig:mz}
\end{figure}

The correlators of $\sigma_x^3$
are much simpler to analyze since they are composed of a fixed number
of Majorana fermions:
\begin{eqnarray}
\label{eq:sigma_Majorana4}
\sigma^3_x&=&A_x B_x\\
\label{eq:sigma_Majorana5}
\sigma^3_x\sigma^3_{x+n}&=&A_x B_x A_{x+n} B_{x+n}.
\end{eqnarray}
Averaging these expressions with the help of
Eq.(\ref{eq:sublattices}), we find
\begin{eqnarray}
\label{eq:m^z}
m^z=\langle\sigma_x^3\rangle&=&1-2e^{-\pi z}I_0(\pi z)\\
\label{eq:corr_m^z}
\langle\sigma_x^3\sigma_{x+2n}^3\rangle&=&\langle\sigma_x^3\rangle^2-
\lp 2e^{-\pi z}I_n(\pi z)\rp^2.
\end{eqnarray}
where $n\ne0$ and, as before, the brackets $\la...\ra$
denote the averages in the full lattice.
The magnetization $m^z$ is plotted in Fig. \ref{fig:mz}.

The behavior of magnetization at fast and slow sweep
is given by the small-$z$ and large-$z$ asymptotic:
\be
\label{eq:transverse_magnetization}
m^z=
\begin{cases}
-1+2 \tilde z+O(z^2)&\tilde z\ll 1\\
%% -\frac{3}{2}\tilde z^2+O(z^3)&\tilde z\ll 1\\
1-(2/\pi \tilde z)^{1/2}+O(z^{-3/2})&\tilde z\gg1
%% -\sqrt{\frac{2}{\pi \tilde z}}-\frac1{\sqrt{32\pi \tilde z^3}}+O(z^{-5/2})&\tilde z\gg1
\end{cases},
\ee
where $\tilde z=\pi z$. 
The asymptotic expansion
in Eq.(\ref{eq:transverse_magnetization}) is in integer powers at small $z$,
and in negative half-integer powers at large $z$. 
The small $z$ limit corresponds to fast sweep,
and so the magnetization deviates little from the initial
$H(t\to-\infty)$ ground state value of $m^z=-1$.
In contrast, large $z$ describe slow sweep when the magnetization 
follows the dynamical field $h(t)$ nearly adiabatically,
and thus $m^z$ approaches the $H(t\to+\infty)$ ground state value 
$m^z=+1$.

The magnetization correlator $\langle\sigma_x^3\sigma_{x+2n}^3\rangle$ is also a smooth function 
of $z$. Subtracting $\langle \sigma_x^3\rangle^2$, 
we obtain the irreducible (connected) correlator
\be
D_n=\langle\sigma_x^3\sigma_{x+2n}^3\rangle-\la \sigma_x^3\ra^2
=-\lp 2e^{-\pi z}I_n(\pi z)\rp^2
. 
\ee
Correlations of magnetization 
at distant points are given by 
the large-$n$ expansion of $D_n$
at fixed $z$. We obtain 
\[
e^{-\pi z}I_n(\pi z)=\int_{-\pi}^\pi e^{\pi z(\cos\theta -1)}
e^{in\theta}\frac{d\theta}{2\pi}
=(2\pi z)^{-1/2}e^{-n^2/2z}
\]
where we used an expansion near the saddle point,
$1-\cos\theta=\frac12\theta^2+O(\theta^4)$, 
and a gaussian approximation for the integral over $\theta$.
This gives an asymptotic behavior
\be
D_n=-\frac{2}{\pi z}e^{-n^2/z}
\ee
The correlation length, which is very short at small $z$ (fast sweep), grows
as $\ell\propto z^{1/2}$ at large $z$ (slow sweep), in agreement with
KZ picture.

\section{Conclusion}
\label{sec:conclusion}

This article presents an exact solution
for a quantum spin chain 
driven through quantum critical points.  
We consider an anisotropic XY chain 
in a time-dependent transverse field $h(t)$ that drives the system from a disordered paramagnetic
phase at early times into an ordered Ising phase, 
and back into the paramagnetic phase 
at late times, 
crossing two quantum critical points along the way.
We construct an exact many-body evolution operator in fermionized
representation with the help of Landau-Zener transition theory,
and use it to study the evolved
state. It is found that the evolved many-body state, while technically
a pure state, acquires local properties of a mixed state.
The emerging nonequilibrium steady state is characterized by finite
entropy density, which is a function of the sweep speed. 
The transformation of a pure state into an entropic state,
resulting from intrinsic decoherence, is analyzed via coarse-graining
in momentum space.

Correlation functions in the final entropic state are calculated using 
the method of Toeplitz determinants. We present exact results
for the 
the asymptotic behavior of spin correlators at large spatial separation.
The correlation length dependence on the sweep speed is found
to be consistent with
the Kibble-Zurek $-1/2$ power law scaling.
We characterize the crossover behavior in which the correlation functions,
monotonic at fast speed, acquire oscillatory spatial dependence
at slow speed. The critical speed for this transition
is found near which the the 
correlation function parameters
exhibit nonanalytic behavior.

\begin{acknowledgements}
R. W. C. acknowledges support from MIT Class of 1995 UROP Fund 
and NDSEG program.
%% completed a portion of this work at MIT with support from MIT UROP.
\end{acknowledgements}

\appendix
\renewcommand{\theequation}{\Alph{section}.\arabic{equation}}

\section{\label{app:toeplitz}Toeplitz Determinant Asymptotics}

Toeplitz matrices, having constant diagonals, 
and their determinants, Eq.(\ref{eq:toeplitz_det1}),
arise in many mathematical and physical problems. 
In particular, one is often interested in
the large-$n$ behavior of $D_n[f]$.
Toeplitz determinant asymptotics form basis for a number of rigorous results,
being particularly useful in the computation of various quantities
in two-dimensional Ising model (see for example Ref. \cite{wu}).
However, the rather daunting mathematical literature on the subject has led to 
some confusion on the status and use of various results such as
Szeg{\"o}'s limit theorem 
and generalizations of the Fisher-Hartwig conjecture
(for example, see Chap. 10 of Ref. \cite{bottcher1}).
We give a formulation of Toeplitz determinant asymptotics that unifies all previously
known results, both conjectured and mathematically rigorous, 
and extends them to a larger class of Toeplitz determinants.

The central quantity in the study of Toepltiz determinants
is a function of complex variable, called generating function, which is
specified for some contour $C$ that encloses
the origin once.
The generating function $f_C(\xi)$ integral over $C$
gives the matrix elements $f_n$ via
\be
\label{eq:fourier_coeff}
f_n=\oint_{C}\frac{d\xi}{\xi}\xi^{-n}f_C(\xi)
\ee
The theory of Toeplitz determinants links the large-$n$ behavior of
$D_n[f]$ to the analytic structure of $f_C(\xi)$, in particular its singularities.
Here we wish to stress two points.  The first point is the importance of specifying the contour
$C$ in relating $f_C(\xi)$ to $f_n$ as it gives an explicit distinction between
singularities inside, outside, and on $C$.  This point is well-known in the
literature where $C$ is taken to be the unit circle and figures prominently in the derivation of 
known results on Toeplitz determinants.

The second point is the freedom to deform $C$ to $C'$ when $f(\xi)$ is analytic between the two contours.
This point was briefly mentioned in Ref. \cite{forrester} and used to obtain the behavior
of spin correlation functions of the two-dimensional Ising model above the transition temperature.
The freedom to deform $C$ in a general setting is a key element of our proposed
extension of Toeplitz asymptotics.

We consider the class of generating functions, first studied by Fisher and Hartwig \cite{fisherhartwig}, 
which are given by
\be
\label{eq:toeplitz_function}
f_C(\xi)=
%% e^{h_0+h_+(\xi)+h_-(\xi)}
e^{H(\xi)}
\xi^m\prod_p
\left(1-\lambda_p^{-1}\xi\right)^{\alpha_p}
\left(1-\lambda_p\xi^{-1}\right)^{\beta_p}
\ee
with 
$H(\xi)=h_+(\xi)+h_-(\xi)+h_0$, where
$h_+$ ($h_-$) is analytic inside and on $C$ (outside and on $C$) satisfying $h_+(0)=0$ ($h_-(\infty)=0$),
and $m$ is an integer winding number.  The roots $\lambda_p$ are on $C$ and give power-law singularities with exponent $\alpha_p$ ($\beta_p$).
However, this representation of $f_C(\xi)$ is not unique in the sense that different choices of 
$C$, $h_0$, $h_\pm$, $\lambda_p$, $\alpha_p$, and $\beta_p$ will give the same matrix elements $f_n$.
For fixed $C$, the formal identity
\be
\label{eq:formal_id}
1=\left(-\xi^{-1}\lambda\right)^n\left(1-\lambda^{-1}\xi\right)^n\left(1-\lambda\xi^{-1}\right)^{-n}
\ee
for integer $n$ shows that the transformation
\begin{eqnarray}
\label{eq:equivalent_rep_first}
e^{h_0}&\rightarrow&e^{h_0}\prod_p(-\lambda_p)^{n_p}\\
m&\rightarrow&m-\sum_p n_p\\
\alpha_p&\rightarrow&\alpha_p+n_p\\
\label{eq:equivalent_rep_last}
\beta_p&\rightarrow&\beta_p-n_p
\end{eqnarray}
gives a generating function with the same Fourier coefficients
$f_n$, Eq.(\ref{eq:fourier_coeff}), as those obtained for the original function $f_C(\xi)$. 
Under such a transformation,
while the parameters $h_0$, $\alpha_p$, $\beta_p$, $m$ 
change, the Toeplitz matrix is preserved.  
The consequences of such transformation 
was first pointed out by Basor and Tracy in \cite{basor3},
who noted that all different generating function
representations contribute to the asymptotics.

Let us now consider deformations of the contour $C$.  
We note that Eq.(\ref{eq:toeplitz_function}) allows singularities to be on $C$.
Since the matrix elements $f_n$ given by Eq.(\ref{eq:fourier_coeff}) must remain the same
upon deforming $C$ to $C'$, such a deformation must not enclose any singularities,
but $C'$ can possibly pass through additional singularities that $C$ does not.
In the representation (\ref{eq:toeplitz_function}),
singularities strictly outside (inside) of $C$ are
described by $h_-$ and $h_+$ while singularities on $C$ are described by the roots $\lambda_p$.  By appropriately
deforming $C$ to $C'$, we can move power-law singularities from $h_+$ and $h_-$ and include them in additional roots 
$\lambda'_p$ on $C'$.

The most general result in the literature is for $C$ fixed to be the unit circle 
but taking into account
the transformations of Eqs.(\ref{eq:equivalent_rep_first})-(\ref{eq:equivalent_rep_last}).  
It was first proposed by Basor and Tracy \cite{basor3}
and is known as the generalized Fisher-Hartwig conjecture.
Each representation given by 
Eqs.(\ref{eq:equivalent_rep_first})-(\ref{eq:equivalent_rep_last})
gives a contribution to $D_n[f]$ of the form
\be
\label{eq:fisher_hartwig}
\delta_{m,0} A e^{h_0 n} n^{\sum_{p}\alpha_p\beta_p} 
\ee
with the prefactor
\begin{widetext}
\be
A=
%% E[h+,h-]
\exp\lp \sum_{k=1}^{\infty}k h_k h_{-k}\rp
\prod_p\frac{G(1+\alpha_p)G(1+\beta_p)}
{G(1+\alpha_p+\beta_p)}
e^{-\alpha_p h_-(\lambda_p)-\beta_p h_+(\lambda_p)}
\prod_{p'\ne p}\left(1-\frac{\lambda_{p'}}{\lambda_{p}}\right)^{-\alpha_{p}\beta_{p'}}
\ee
\end{widetext}
where $G(x)$ is the Barnes $G$-function \cite{basor3} which satisfies $G(x+1)=\Gamma(x)G(x)$ and 
%% $\ln E[b+,b-]=\sum_{k=1}^{\infty}k h_k h_{-k}$ where 
%
\[
h_{\pm}(\xi)=\sum_{k=1}h_{\pm k}\xi^{\pm k}
.
\]
The constraint $\delta_{m,0}$ in Eq.(\ref{eq:fisher_hartwig})
means that the contributions for non-zero 
winding numbers $m$ 
are not of the above form but decay faster than $n^\eta$ for all real $\eta<0$.
The asymptotic of $D_n[f]$ is obtained by summing 
the terms which give the leading contribution for large $n$.

This conjecture has been proven rigorously in some cases. 
The case for arbitrary $\alpha_p$ and $\beta_p$, but 
such that only one representation contributes to the leading 
term, has only been proven relatively recently \cite{ehrhardt1,ehrhardt2}.  
The case for positive integer $\alpha_p$ and $\beta_p$
but with multiple representations contributing at leading order was proven by B{\"o}ttcher and Silbermann \cite{bottcher2}.

The generalized Fisher-Hartwig conjecture as stated above 
gives the asymptotics of $D_n[f]$ as a sum of
contributions from each equivalent representation 
of the generating function $f_C(\xi)$, 
but with $C$ fixed to be the unit circle.  
The natural extension is to also allow arbitrary deformations of $C$ to $C'$ that 
may touch
but not cross the singularities, and then sum over the leading contributions from the additional 
equivalent representations.  
This procedure can be concisely expressed as follows. One writes
down the generating function in the form of Eq.(\ref{eq:toeplitz_function}) with the power-law 
singularities for $\lambda_p$ arbitrarily distributed in the complex plane.
Then one generates the equivalent representations via 
Eqs.(\ref{eq:equivalent_rep_first})-(\ref{eq:equivalent_rep_last})
and sums
over the contributions given by Eq.(\ref{eq:fisher_hartwig}).
In practice only the singularities closest to the unit circle need to be considered.
This is because under the transformation of 
Eqs.(\ref{eq:equivalent_rep_first})-(\ref{eq:equivalent_rep_last}), $e^{h_0}$ 
gets multiplied by powers of $\lambda_p$.  Since the contribution to $D_n[f]$ given by 
Eq.(\ref{eq:fisher_hartwig}) contains $e^{h_0 n}$, the singularities far from the unit circle
generically give subleading contributions.

This proposed extension of the generalized Fisher-Hartwig conjecture is particularly useful
for generating functions with power-law singularities off of the unit circle but non-zero winding number.
The generalized Fisher-Hartwig conjecture is not applicable in this case due to the presence of
winding numbers. [Note $\delta_{m0}$ in Eq.(\ref{eq:fisher_hartwig}).]
However, by using the freedom to
deform the contour $C$ to the singularities
off of the unit circle we can absorb the winding number into $\alpha_p$, $\beta_p$ via
Eqs.(\ref{eq:equivalent_rep_first})-(\ref{eq:equivalent_rep_last}).
After that the zero winding number result, Eq.(\ref{eq:fisher_hartwig}),
can be used. 

Essentially, the generalized Fisher-Hartwig conjecture relates power-law singularities on the unit circle
to the asymptotic behavior of Toeplitz determinants.  
Our proposed extension just states that the same relation 
holds for power-law singularities with a generic location 
in the complex plane.  
The literature on Toeplitz determinants
mostly considers singularities on the unit circle, 
with an exception of the result obtained by Day
\cite{day} 
for rational generating functions
\be
\label{eq:days_function}
f(\xi)=\frac{\prod_p(\xi-\lambda_p)}{\prod_q(\xi-\rho_q)}
\ee
where $\lambda_p$ and $\rho_q$ are arbitrary in the complex plane.  This 
function is clearly of the form
of Eq.(\ref{eq:toeplitz_function}).  The corresponding Toeplitz determinant can be evaluated
explicitly and the result is given exactly by generating all equivalent representations using
all the roots via 
Eqs.(\ref{eq:equivalent_rep_first})-(\ref{eq:equivalent_rep_last})
and summing over the corresponding contributions of 
Eq.(\ref{eq:fisher_hartwig}).  
This provides evidence that the extension proposed here holds in a general setting, although
we expect it to give
%% reliably give 
only the leading asymptotic contribution and not the exact determinant in this case.

\bibliography{paper}
\bibliographystyle{apsrev}

\end{document}